**The Neoplasia as embryological phenomenon and its implication in the animal evolution and the origin of cancer. II. The neoplastic process as an evolutionary engine**

Jaime Cofre

Laboratório de Embriologia Molecular e Câncer, Federal University of Santa Catarina, room 313b, Florianópolis, SC, 88040-900, Brazil

Corresponding author. Laboratório de Embriologia Molecular e Câncer, Universidade Federal de Santa Catarina, Sala 313b, Florianópolis, SC, 88040-900, Brazil

E-mail address: jaime.cofre@ufsc.br

**Abstract**

In this article, I put forward the idea that the neoplastic process (NP) has deep evolutionary roots and make specific predictions about the connection between cancer and the formation of the first embryo, which allowed for the evolutionary radiation of metazoans. My main hypothesis is that the NP is at the heart of cellular mechanisms responsible for animal morphogenesis and, given its embryological basis, also at the center of animal evolution. It is thus understood that NP-associated mechanisms are deeply rooted in evolutionary history and tied to the formation of the first animal embryo. In my consideration of these arguments, I expound on how cancer biology is perfectly intertwined with evolutionary biology. I describe essential cellular components of unicellular holozoans that served as a basis for the formation of the neoplastic functional module (NFM) and its subsequent exaptation, which brought forth two great biophysical revolutions within the first embryo. Finally, I examine the role of Physics in the modeling of the NFM and its contribution to morphogenesis to reveal the totipotency of the zygote.

**The role of cadherins: cells learning to proliferate together toward embryonic multicellularity**

From the moment of fertilization or the beginning of embryogenesis onward, there are remarkable similarities between cancer and embryo cells. Cell division begins with a single cell, the zygote, and consolidates during the segmentation stage, establishing a group of cells that, I here hypothesize, have "learned" to stay together during initial embryogenesis through the engagement of cadherins or similar molecules expressed on the cell surface (Harwood and Coates, 2004). The connection between cadherins and cancer is well established in the literature (Mendonsa, Na and Gumbiner, 2018; Yu *et al.*, 2019; Hegazy *et al.*, 2022). Likewise, studies confirmed the expression of cadherins in protists (Y.-P. Chen *et al.*, 2019) and unicellular holozoans (King, Hittinger and Carroll, 2003; Sebé-Pedrós and Ruiz-Trillo, 2010; Nichols *et al.*, 2012; Suga *et al.*, 2013; Hehenberger *et al.*, 2017). Thus, based on these observations, I propose an early co-option of cell proliferation and adhesion mechanisms, linked to an initial provocation and induced by the fusion of protist reproductive cells (fertilization).

Some characteristics of ctenophores from the order Beroida are particularly relevant to this discussion because they serve as illustrative examples of transient adherens junctions (Tamm and Tamm, 1991). Transient intercellular bridges in the buccal epithelium of ctenophores are associated with the actin cytoskeleton and represent the fastest events of junction derangement/disappearance observed so far (Tamm and Tamm, 1993). Such a characteristic provides buccal cells with different possibilities of reorganization, a behavior comparable to that of tumor cells during the growth phase or embryonic cells in motion (Tamm and Tamm, 1993). These cells are able to stick together and separate quickly. Both separation and mobility are important for the expansion of the oral cavity and, therefore, essential to the predatory behavior of ctenophores. In my view, this mechanism could stand for one of the first steps in the formation of the first embryo, whereby cells "learned" to stay connected without losing mobility, which would later be required for morphogenesis, organogenesis, and tissue reorganization in the animal itself.

Some observations inspiringly support my hypothesis of cells "learning to stay together," such as the cellularization stage of *Creolimax fragrantissima* (Sebé-Pedrós, Degnan and Ruiz-Trillo, 2017), the process of colony formation of *Capsaspora owczarzaki* (Sebé-Pedrós, Irimia, *et al.*, 2013), *Ministeria vibrans* (Mylnikov *et al.*, 2019), *Pigoraptor vietnamica*, and *Pigoraptor chileana* (Hehenberger *et al.*, 2017), and segregation and dispersal movements of amoeboid and filopodial amoeboid cells. These observations suggest that the cellular basis of unicellular holozoans is able to maintain cohesion through the extracellular matrix (ECM) (Sebé-Pedrós, Irimia, *et al.*, 2013), interacting via filopodia (Mylnikov *et al.*, 2019) or forming some type of syncytium-derived epithelium (Sebé-Pedrós, Degnan and Ruiz-Trillo, 2017), while retaining the ability to disperse at appropriate times (Marshall *et al.*, 2008; Suga and Ruiz-Trillo, 2013, 2015). Thus, the capacity to collaborate and stick together without impairing movability is imperative during the life cycle of unicellular holozoans, and a requisite condition for the emergence of metazoans. In line with this proposition, I speculate that the first step in the formation of the first embryo would involve a type of "benign tumor" of multicellular clonal origin, constituting a collaboration of closely associated cells with the potential to migrate at opportune moments.

**Actin cytoskeleton: a first draft of morphogenesis in unicellular and multicellular colonies**

Ubiquitous elements of animal cells such as actin and myosin are also found in protists (Sebé-Pedrós, Burkhardt, *et al.*, 2013). Interestingly, this was a topic of controversy among scientists for several decades (Grain, 1986; Dawson and Paredez, 2013), as only the existence of well-structured microtubules was first widely accepted (Grain, 1986). Later, actin was found to be expressed even in the cell nucleus of protists (Soyer, 1981; Berdieva *et al.*, 2016). The presence of true microvilli (i.e., permanent projections supported by axial microfilament bundles) was confirmed in Filasterea (Sebé-Pedrós, Burkhardt, *et al.*, 2013; Hehenberger *et al.*, 2017) and Choanoflagellatea (Sebé-Pedrós, Burkhardt, *et al.*, 2013). A microfilament-organizing center (MFOC) was ultrastructurally identified in *M. vibrans* (Mylnikov *et al.*, 2019), where it seems to act as a source of F-actin for microvilli, sharing morphological similarities with the axoplast (an MFOC that organizes microtubular axonemes of the axopodia), which is well documented in Heliozoa and Radiolaria (Anderson, 1983; Mylnikov *et al.*, 2019). The presence of an MFOC in the cell center of *M. vibrans* (Mylnikov *et al.*, 2019), as well as the existence of an extensive skeleton of microfilaments (Shalchian-Tabrizi *et al.*, 2008), suggests a preponderant role of the actin cytoskeleton in unicellular Holozoa. It is clear that the general arrangement of actin has been conserved throughout evolution, and there is ample evidence of the expression of different types of actin, encoded by different gene families, in protists (Yi *et al.*, 2016).

Still within the context of actin microfilaments, another noteworthy characteristic of protists is exemplified by colony formation in multicellular green algae of the genus *Volvox*. *Volvox* colonies are held together predominantly by cytoplasmic bridges produced during the cleavage of the asexual embryo (Green and Kirk, 1981). Patterning of the body plan is achieved using a stereotypical "development program" including embryonic cleavage with asymmetric cell division, morphogenesis (i.e., inversion, also known as gastrulation), and cell differentiation (Matt and Umen, 2016). Initially, there were limited data supporting the participation of the actin cytoskeleton in the gastrulation phase of *Volvox*, but actomyosin contraction is currently believed to be essential for colony inversion (Nishii and Ogihara, 1999; Matt and Umen, 2016). A phenomenon similar to inversion was observed in choanoflagellates, whereby several cells act together to produce an internal mechanical force dependent on actomyosin contractility (Brunet *et al.*, 2019). This collective contraction would share similar evolutionary roots with the actomyosin contractile ring observed during zebrafish epiboly (Behrndt *et al.*, 2012; Hernández-Vega *et al.*, 2017). Mechanical force-dependent cellularization and actomyosin contractility were also reported in Ichthyosporea (Dudin *et al.*, 2019). At any rate, what is relevant to my hypothesis is the multipotentiality of protists and their great reliance on the actin cytoskeleton, given its organization capacity, participation in the cell nucleus, and pivotal roles in the morphogenesis of colonial organisms as well as in the life cycle of species with temporary stages of multicellularity.

In examining this topic, I must be careful to separate the activities of the actin cytoskeleton and its associated proteins (e.g., cell support) from the activities of mechanical tension (Wozniak and Chen, 2009) and mechanical memory systems (Balestrini *et al.*, 2012; Heo *et al.*, 2015; Li *et al.*, 2017), given that only the latter are, in my understanding, truly emergent properties (Stephan, 1998). As I intend to formulate my hypothesis for the origin of the first embryo based on Neoplasia, it is important to clarify that, if out of control, such processes would inevitably lead to the collapse or dispersion of the embryo, as supported by observations on *C. fragrantissima* and *C. owczarzaki* colonies. Therefore, in the embryo self-organization model, the existence of a Neoplasia control system is implicit, which is in agreement with the previously proposed concept of cells "learning" to grow together. Cadherins represent one of the cellular systems that could help control cell proliferation to a sufficient extent. Nevertheless, several other mechanisms warrant investigation for their associations with multicellular organization, such as integrins, the Hippo pathway, and the ECM, all of which are well characterized in unicellular holozoans. These mechanisms will be described below.

**Integrin adhesome**

Integrins are signal transduction molecules anchored in the cell membrane that associate with a group of intracellular proteins able to interact with the actin cytoskeleton, forming what is known as the integrin-mediated adhesive complex (IMAC). Integrins and the IMAC have critical roles in cell development, migration, and proliferation. Our understanding of the origin of animals changed drastically when integrins were detected in unicellular Holozoa, demonstrating that many features previously thought to be animal-specific (Suga *et al.*, 2013) were actually passed down to Metazoa by their unicellular progenitors. Integrins and near-complete components of the IMAC were reported in *C. owczarzaki* (Sebé-Pedrós *et al.*, 2010), *Pigoraptor* spp. (Hehenberger *et al.*, 2017), *Syssomonas multiformis* (Hehenberger *et al.*, 2017), *Corallochytrium limacisporum* (Grau-Bové *et al.*, 2017), *C. fragrantissima* (De Mendoza *et al.*, 2015), *Sphaeroforma arctica* (Dudin *et al.*, 2019), *Thecamonas trahens* (Sebé-Pedrós *et al.*, 2010), *Pygsuia biforma* (Brown *et al.*, 2013), and *M. vibrans* (Kang *et al.*, 2020).

In *P. chileana*, the presence of fibronectin-3 domains in several receptor tyrosine kinases and a receptor tyrosine phosphatase (Hehenberger *et al.*, 2017), known to interact with integrins (Pankov and Yamada, 2002), suggests a possible association between the integrin adhesome and tyrosine kinase/phosphatase signaling. A similar association was proposed for *C. owczarzaki*, also shown to contain receptor tyrosine kinases with fibronectin-3 domains (Sebé-Pedrós, Irimia, *et al.*, 2013). Another example, the choanoflagellate

*Monosiga brevicollis*, harbors a wide range of tyrosine kinase proteins, including Ras, Rho, Rac, Cdc42, small GTPases, PLC (which functions as a PI3K/Ca$^{2+}$ signaling adapter), and SRC subgroup tyrosine kinases (Manning *et al.*, 2008). Outside the Metazoa, a focal adhesion kinase (FAK) was described only in *C. owczarzaki* and *M. vibrans* (Sebé-Pedrós *et al.*, 2010; Suga *et al.*, 2012). It is also noteworthy the transient cellularization observed in some stages of the life cycle of unicellular Holozoa, leading to the formation of a polarized epithelium expressing α- and β-integrin receptors as well as α- and β-catenins (Dudin *et al.*, 2019). These findings demonstrate the importance of cell–matrix and cell–cell adhesion bonds in the life cycle of unicellular Holozoa, and their relevance to my proposal of a neoplastic functional module (NFM).

**Hippo signaling pathway**

Some of the essential aspects of my hypothesis are the co-option of the Hippo pathway and the elucidation of its phylogenetic origin. The Hippo pathway was first described in *Drosophila melanogaster* (Justice *et al.*, 1995; Xu *et al.*, 1995). Hippo mutations resulted in exaggerated growth in a variety of epithelial structures, such as wings, legs, and eyes. Interestingly, although Hippo-mutant clones were able to correctly produce the monolayer organization that is typical of imaginal discs, these cells exhibited abnormalities on the apical surface, forming a projection of the cell body that gave them a dome-like appearance (Justice *et al.*, 1995). More recent studies revealed a conserved function of this pathway in the control of organ size in mammals (for a complete review of Hippo pathway components, see Pan, 2007; Badouel et al., 2009).

There is convincing evidence that an active Hippo signaling pathway was already present in the unicellular ancestors of Metazoa. Accordingly, Yki orthologs were identified in

*C. owczarzaki*, *M. brevicollis*, and *Salpingoeca rosetta* (Sebé-Pedrós *et al.*, 2012). Non-metazoan orthologs of Yki contain highly conserved functional sites, such as the Hippo pathway-responsive phosphorylation site S168/127 and the N-terminal homology region, the latter of which is critical for interaction with the Scalloped (Sd)/TEAD transcription factor. Hippo orthologs were identified in *C. owczarzaki* and *S. rosetta* by the presence of a Ste20-like kinase domain and a SARAH domain (Sebé-Pedrós *et al.*, 2012). Among non-metazoans, Wts and Sd proteins were found only in *C. owczarzaki*. This species was also found to express several upstream regulators of Hippo, such as Kibra, Mer, aPKC, and Lgl (Sebé-Pedrós *et al.*, 2012). Therefore, the level of regulatory complexity of the Hippo/YAP pathway in Filasterea is extremely high. In mammals, only Kibra and Mer were identified (Pan, 2007), indicating a significant reduction of functional proteins of the Hippo pathway from filastereans to humans.

Hippo signaling is known to be activated in a cell density-dependent manner (see review by Zhao et al., 2010; Pan, 2010) through a process that, in cell culture, is mediated by the Mer regulator (McClatchey and Giovannini, 2005; Okada, Lopez-Lago and Giancotti, 2005). Mer is involved in reciprocal signaling with the main effectors of the integrin adhesome, constituting a genuine interdependence between cell adhesion and cell cycle progression (Pugacheva, Roegiers and Golemis, 2006).

It should be noted, however, that the function of the Hippo pathway has not always been to regulate cell proliferation. In *C. owczarzaki*, Hippo is believed to regulate actin cytoskeleton dynamics and colony morphogenesis (Phillips *et al.*, 2021). Some of these old functions seem to have been retained throughout evolution, given that YAP, TAZ, and Yorkie participate in actin cytoskeleton-dependent mechanotransduction systems (Dupont *et al.*, 2011; Driscoll *et al.*, 2015) and actin-dependent mechanical memory, which influence cell differentiation (Yang *et al.*, 2014). I would like to emphasize that Hippo, currently considered a tumor suppressor, is likely to have participated in the "learning" process that allowed cells to proliferate together. I also underscore the importance of the recruitment of these components into the NFM for the construction of an organized animal embryo.

Interestingly, *C. owczarzaki* is so far the only non-metazoan known to harbor all components of the integrin-mediated signaling and adhesion system (Sebé-Pedrós *et al.*, 2010). Members of the genus *Capsaspora* are also known to carry genes related to the Hippo pathway, such as Myc (Sebé-Pedrós *et al.*, 2011) and cyclin E (Sebé-Pedrós *et al.*, 2012). The relationship of Hippo pathways with cancer has been addressed in several reviews, and there is no doubt about the importance of these pathways in tumorigenesis (Pan, 2010; Zhao *et al.*, 2010).

**The ECM as a requirement for multicellularity**

The ECM not only serves as a selective barrier to macromolecules, provides support to epithelial cells, and regulates cell migration (Morrissey and Sherwood, 2015) but also is considered an essential element in the transition to multicellularity and tissue evolution (Özbek *et al.*, 2010; Fidler *et al.*, 2018). A study investigating early animals found that ctenophores contain collagen IV and that the number and diversity of collagen IV genes in these organisms exceed that of any other group of metazoans (Fidler *et al.*, 2017). In the referred study, transcriptome analysis was used to compare 10 ctenophores, namely *Mnemiopsis leidyi*, *Pleurobrachia pileus*, *Pleurobrachia bachei*, *Beroe ovata*, *Beroe abyssicola*, *Euplokamis* sp., *Dryodora* sp., *Vallicula* sp., *Coeloplana* sp., and *Bolinopsis* sp. Altogether, 118 unique collagen IV genes were detected. Each organism contained a variable number of collagen IV genes, ranging from 4 to 20, a value much higher than those found in other animal phyla (2ó6 genes) (Fidler *et al.*, 2017).

Concerning unicellular Holozoa, which are considered the closest relatives of animals, a study reported the discovery of a collagen IV-like gene in *M. vibrans* (Grau-Bové *et al.*, 2017). This finding indicates that collagen IV has pre-metazoan ancestry and serves some function in individual cells, suggesting a potential role of collagen IV in the transition from unicellular organisms to multicellular animals. However, in a recent study, the collagen region of *M. vibrans* was found to be very short (81 Gly-XY repeats), differing greatly in size from the collagen IV region of metazoans (that of *Mus musculus*, for example, contains 443 repeats) (Linden and King, 2021). The same study also reported that 22 choanoflagellate species, 3 filasterean species (*M. vibrans*, *P. vietnamica*, and *P. chileana*), and 1 ichthyosporean species (*S. arctica*) encode proteins containing collagen domains for secretion into the ECM (Linden and King, 2021).

These differences in the number of collagen IV repeats between *M. vibrans* and metazoans do not represent a problem from the point of view of my hypothesis. The exon theory of genes, based on the intron/exon structure of eukaryotic genes, posits that new genes evolve through exon shuffling, with introns being fundamental to evolutionary processes (Gilbert, 1987). Regardless of the origin of introns in ancient genes, exon shuffling provides an explanation for the formation of new functional proteins (Patthy, 1987) through a process mediated by intronic recombination (Patthy, 1999). Exon shuffling is believed to occur between introns of the same phase or via insertion of a symmetric exon into an intron of the same phase, thereby ensuring phase compatibility in the resulting coding sequence. According to László Patthy, exon shuffling acquired great significance at the time of metazoan radiation, a phenomenon coinciding with a spectacular explosion of evolutionary creativity (Patthy, 1999).

In agreement with this idea, the rate of creation of multi-domain proteins accelerated in the metazoan lineage (Tordai *et al.*, 2005), which can be partially explained by the frequent insertion of exon-bordering domains into novel protein architectures (Ekman, Björklund and Elofsson, 2007). This might have happened in a context of proliferation and co-option of the mechanisms of Neoplasia implied in the formation of the first embryo. I purport that the neoplastic process (NP) facilitated intronic recombination in the metazoan lineage and speculate that such a process contributed significantly to the emergence of new complex multi-domain proteins, new functions, and increased organismic complexity in metazoans.

Finally, as is widely known, scientific evidence shows that genomic rearrangements are characteristic of tumor cells. The study of genetic rearrangements in humans can be performed by using Alu repetitive elements (Elliott, Richardson and Jasin, 2005), which constitute the largest family of repeats, reaching about one million copies and representing 11% of the human genome (Lander *et al.*, 2001). In somatic cells, recombination between intronic Alu elements was shown to lead to partial duplication of the *MLL* (*ALL1*) gene in patients with acute myeloid leukemia (AML) (Caligiuri *et al.*, 1994). *ALL1* gene rearrangement was also detected in 2 patients with trisomy 11 (Schichman *et al.*, 1994) and 10 patients (18% of the sample) with M4-/M5-subtype AML (So *et al.*, 1997). Partial tandem duplication of the *ALL1* gene was detected in 11% of patients with AML and normal cytogenetics (Strout *et al.*, 1998). Furthermore, genomic rearrangements of the *BRCA1* gene were identified in families with a history of breast and/or ovarian cancer (Puget *et al.*, 2002), and aberrant splicing of APC exon 14 was associated with familial adenomatous polyposis, a syndrome predisposing to autosomal dominant colorectal cancer (Tuohy *et al.*, 2010).

It can be seen that all essential elements of my hypothesis were already present in unicellular holozoans. I propose that these cellular elements were recruited into an NFM, as they are now directly related to cancer. A first exaptation within the NFM module was necessary to produce the great revolution of multicellular collective movements.

**Co-option of basic NP elements and the great animal revolution: the epibolic process**

For cells to be able to grow and proliferate together as an embryo, it was necessary to integrate membrane adhesion proteins (cadherins from adherens junctions) and actin filaments. Hence the functional relevance of a cytoskeleton interlinking this multicellular organization. Of note, mechanical stress generated by cell proliferation seems to play an important role in epithelial growth (Hufnagel *et al.*, 2007; Aegerter-Wilmsen *et al.*, 2010). In embryonic epithelia, such as those initiating the epiboly, where little to no ECM is present (Latimer and Jessen, 2010), mechanical tension and stress produced by actomyosin contraction are transmitted over long distances through adhesion proteins (Charras and Yap, 2018). Also, epithelial delamination during zebrafish epiboly is known to be mediated by the orientation of cell divisions produced by mechanical tension, allowing cells to efficiently release anisotropic tension as the epithelium expands (Campinho *et al.*, 2013). In non-embryonic models, such as monolayer cultures of endothelial and epithelial cells, collective migration is governed by a simple and unifying principle: neighboring cells join forces to transmit a significant amount of normal stress through E-cadherins and migrate so as to produce minimal intercellular shear stress (Tambe *et al.*, 2011).

Thus, for epiboly to occur, it is necessary to establish an integrated cytoskeleton network connected by E-cadherins. I believe that multicellular actomyosin cables (Wood *et al.*, 2002; Franke, Montague and Kiehart, 2005; Fernandez-Gonzalez *et al.*, 2009) constructed during early embryogenesis might have been transcendental to embryonic morphogenesis. Compatible with this idea is the report of actomyosin cables in unicellular protists (Soyer, 1981) and ctenophores (Tamm and Tamm, 1988). It is therefore possible to imagine multicellular actin cables and a continuous and integrated tension network in the first animal embryo, which could use this architecture for self-organization. Such a model has been predicted by the cellular tensegrity theory proposed by Donald Ingber (Ingber, Madri and Jamieson, 1981; Ingber, 1993). Ingber's theory applies the concept of cellular tensegrity to explain mechanochemical transduction (Wang, Butler and Ingber, 1993) and morphogenetic regulation (Ingber, 1993). Ingber himself suggested that integrins may act as mechanoreceptors and may transmit mechanical signals to the cytoskeleton. In my proposal, i add the preponderant role of adherens junctions (E-cadherins) in mechanotransduction during epiboly. This mechanoreceptor function has been well characterized in the last decade (le Duc *et al.*, 2010; Yonemura *et al.*, 2010; Charras and Yap, 2018; Arbore *et al.*, 2020; Bonfim-Melo *et al.*, 2022).

Within a framework that unites physics and embryogenesis, there is a growing body of evidence supporting that contractile forces generated internally by the actomyosin cytoskeleton can act as regulators of cell behavior. Such observations suggest a broader role for mechanotransduction (Wozniak and Chen, 2009), with its participation being important in multiple stages of embryogenesis, including cell fate, growth, morphogenesis, and organogenesis (Lauffenburger and Horwitz, 1996; DuFort, Paszek and Weaver, 2011; Heller and Fuchs, 2015; Morita *et al.*, 2017; Agarwal and Zaidel-Bar, 2021). Tissue-scale morphogenetic movements, such as epiboly, are driven by dynamic remodeling of cell–cell adhesion at cellular interfaces (Lecuit, Lenne and Munro, 2011; Heisenberg and Bellaïche, 2013; Yap, Duszyc and Viasnoff, 2018). Therefore, the expansion of an epithelial monolayer, similar to the transformation seen during epiboly in ctenophores, is being increasingly understood as a mechanical phenomenon (Morita *et al.*, 2017) (Figure 1). It is now known that, in an expanding epithelium, each cell generates forces on the underlying substrate (du Roure *et al.*, 2005). Under these conditions, movement direction and coordination are driven by physical forces (Trepat *et al.*, 2009; Tambe *et al.*, 2014; Zaritsky *et al.*, 2014) and induce the spontaneous generation of mechanical waves (Notbohm *et al.*, 2016; Pajic-Lijakovic and Milivojevic, 2020; Petrolli *et al.*, 2021). The epithelium can transmit forces through intercellular junctions in a manner that creates long-range stress gradients (Serra-Picamal *et al.*, 2012) similar to morphogens. Only recently have studies begun to integrate embryonic mechanics and pattern formation (Heller and Fuchs, 2015).

As it could not be otherwise in the conceptual framework of a hypothesis supporting Neoplasia as an evolutionary engine, physical and mechanical properties prevail in the evolution of cancer (Wirtz, Konstantopoulos and Searson, 2011; Jain, Martin and Stylianopoulos, 2014). Tumor progression is driven, among other characteristics that i will address later in the article, by expansion of the growing tumor mass and increased contractility of tumor cells (Northcott *et al.*, 2018), features that are homologous to the first stages of embryonic development. It has recently been revealed, using biophysical models of three-dimensional (3D) collective cell migration, that mechanical waves are a general characteristic of cells migrating together (Pajic-Lijakovic and Milivojevic, 2022). Collective migration of a human carcinoma cell line (HCT116) produces a large-scale viscoelastic force that influences cell rearrangement and induces the generation of mechanical waves (Pajic-Lijakovic and Milivojevic, 2022). Thus, in the model proposed in this article, mechanical waves would represent a consequence of long-term cellular rearrangement driven by viscoelastic and surface tension forces. Viscoelastic forces are resistive forces always opposite to the direction of migration (Pajic-Lijakovic and Milivojevic, 2020). Moving back from the cancer model to my embryonic model of epiboly, it can be said that the viscoelastic force would be in the aboral direction of the ctenophore, whereas the mechanical wave would be directed toward the oral pole.

Overall, the available evidence indicates that mechanical (physical) forces generated by living cells at the molecular level can propagate to cellular and embryonic levels and have profound implications for pattern formation (Das *et al.*, 2019). This is relevant for modern embryology because it suggests that mechanical forces could have a similar role to morphogens (Agarwal and Zaidel-Bar, 2021). Renowned embryologists such as Mark Martindale have dedicated their efforts to understanding the bases of polarity in ctenophores. Martindale's studies on Wnt showed that the protein is expressed much later than expected and therefore does not participate in initial polarity formation, leaving it a mystery as to how the ctenophore embryo can organize itself at the beginning of embryogenesis without expressing a gene that has such a great influence on body patterning in metazoans (Pang *et al.*, 2010). Other authors, despite detecting Wnt expression in the oral region of *P. pileus*, were not able to demonstrate its role in oral–aboral polarity (Jager *et al.*, 2013).

Delving further into the importance of tension networks in the context of the formation of the first embryo, a recent study showed that mechanical tension regulates the Hippo pathway in *Drosophila* (Fletcher *et al.*, 2018). This investigation demonstrated that mechanical deformation forces increase the nuclear localization of active Yki, thereby promoting the expression of apical proteins. On the other hand, when columnar epithelial cells are densely packed, Yki localizes to the cell cytoplasm. A mechanotransduction system involving Hippo could thus participate in cell proliferation, morphogenesis (Fletcher *et al.*, 2018), and mechanical memory (Yang *et al.*, 2014). Note that the original function of the Hippo pathway in unicellular holozoans, which was apparently conserved in other animals, was to regulate the morphogenesis and dynamics of the actin cytoskeleton (Phillips *et al.*, 2021). The results of Fletcher and colleagues could be applicable to other flat cell epithelia such as the zebrafish blastoderm, wherein radial intercalation occurs during epiboly (Morita *et al.*, 2017). In the context of my hypothesis, these findings indicate the potential for a perfectly harmonic relationship between changes in cell shape, proliferation, and cell movements leading to the success of embryonic morphogenesis.

Thus, co-option of the actin cytoskeleton and its associated proteins, E-cadherins (Gumbiner, 1996), and elements regulating intense cell proliferation might have been crucial to the emergence of epiboly. The third oblique division of ctenophores is relevant in this regard (Pianka, 1974), as it forms a set of micromeres that create an initial circular halo in the aboral region (Martindale and Henry, 1997, 1999, 2015; Henry and Martindale, 2001). In sea urchins, asynchronous cell divisions and the 5th tangential division produce a more squamous morphology in cells forming the blastocele during embryonic development (Wolpert and Gustafson, 1961). The squamous shape of the presumptive enveloping layer (pre-EVL) in *Danio rerio*, which is equivalent to the shape of cells initiating epiboly in ctenophores, depends on force generation by actomyosin networks and on the transmission of these forces through adhesion complexes (Xiong *et al.*, 2014). Collective contraction (similar to a circular ring or halo) and actomyosin-dependent force generation have already been observed in ichthyosporean and choanoflagellate representatives of unicellular holozoans during stages of transient multicellularity (Brunet *et al.*, 2019; Dudin *et al.*, 2019). Therefore, tissue geometry, cell division, and mechanical force interact to produce morphogenesis (Lecuit and Lenne, 2007; Heisenberg and Bellaïche, 2013). Particularly, cell behavior is a function of the ratio of adhesion to cortical tension (contractility) (Manning *et al.*, 2010; Maître *et al.*, 2012). I therefore speculate that this halo or circular ring is a tension network created by multicellular actomyosin cables in the aboral region of the embryo and constitutes a decisive structure for triggering epiboly. As previously mentioned, it is known that the presence of multicellular actomyosin cables has implications for morphogenesis (Wood *et al.*, 2002; Franke, Montague and Kiehart, 2005; Fernandez-Gonzalez *et al.*, 2009) and that there are reports of collective contractility in our closest metazoan relatives (Brunet *et al.*, 2019; Dudin *et al.*, 2019). Zebrafish epiboly, for example, is driven by a contractile actomyosin ring in the external yolk syncytial layer in direct opposition to the blastoderm (EVL) via a mechanism either dependent (Behrndt *et al.*, 2012) or independent (Hernández-Vega *et al.*, 2017) of flow friction.

It is also known that epithelial cells migrate collectively as continuous sheets of multiple cell lines that extend cryptic lamellipodia to collectively control movement (Farooqui and Fenteany, 2005). Thus, i speculate that, when this halo of cells projected onto macromeres, they kick-started the great animal revolution, creating an integrated cytoskeleton network that drives epiboly in the direction of the oral pole (Figure 1). By analyzing the fate of $e_1$ and $m_1$ micromeres in the study of Mark Martindale, it is possible to see a clear link between cells in the aboral-oral direction (see Figures 9D, 4B, 4D, 3B, 3F in Martindale and Henry, 1999). Some authors interpret that, during epiboly, micromeres destined to form the comb rows divide and migrate faster than the rest, creating four thick regions of small cells on the embryo surface (Pianka, 1974). Organization into a tension network would be coherent with the fate of $e_1$ micromeres, which form a wide network interconnected across the entire length of the epidermis (Martindale and Henry, 1999).

This phenomenon of first epiboly, which i call the first great revolution, has immense consequences for the embryological process and biological evolution. I propose that ectodermal epithelial cells would have migrated together according to a type of collective migration with precedents in evolutionary history. Amoeboid cells of *Dictyostelium discoideum* are highly mobile, with dynamic and short-lived filopodia (Medalia *et al.*, 2007; Arthur *et al.*, 2021). These organisms come together to form a moving mass of up to $10^5$ cells, which collectively migrate over a substrate (Palsson and Othmer, 2000). Some of the molecular mechanisms involved in collective migration have been elucidated (Mathavarajah *et al.*, 2021). It seems plausible to think that some characteristics of collective migration observed in embryos were acquired from the protist ancestor of animals. In opisthokonts, which are placed a little closer to the origin of metazoans, the amoeba *Fonticula alba* is reported to perform collective movements (Toret *et al.*, 2022). In filastereans, there are reports of a filopodial amoeboid stage in *C. owczarzaki* (Sebé-Pedrós, Irimia, *et al.*, 2013).

Although there is no evidence that unicellular holozoan cells associate to migrate collectively, these organisms are well-known to exhibit transient multicellularity, cellularization, and formation of a polarized epithelium dependent on actin cytoskeleton contractility, as reported in the ichthyosporean *S. arctica* (Dudin *et al.*, 2019). Also in *S. arctica*, a prototype of an actomyosin contractile ring and a transient modular association (temporal co-expression) of cytoskeleton proteins (formin, Arp2/3, septins, cofilin, profilin, myosin II, and myosin V) were found to be expressed concomitantly with cell adhesion proteins ( -and  -integrin receptors and  -and  -catenin) at a specific stage of cellularization (Dudin *et al.*, 2019). Another example of collective cell contractions was observed in the choanoflagellate *Choanoeca flexa*, which temporarily associates and produces actomyosin-mediated apical contractility (Brunet *et al.*, 2019). These findings suggest that the last common ancestor of holozoans was an organism capable of transient multicellularity whose cells could contract collectively, produce an internal mechanical force, and establish a functional module for co-expression of adhesion proteins and cytoskeletal proteins, concordant with my proposal of an NP involved in the first epibolic revolution. Thus, the first animal embryos could be pictured as amoeboid cells or filopodial amoeboids migrating collaboratively by applying mechanisms acquired in unicellular Holozoa (Sebé-Pedrós, Irimia, *et al.*, 2013; Hehenberger *et al.*, 2017; Sebé-Pedrós, Degnan and Ruiz-Trillo, 2017; Mylnikov *et al.*, 2019).

This proposal of an NP at the heart of animal evolution carries implications not only for the embryological process, as described earlier, but also for cancer. Several types of cancer cell migration are currently recognized: amoeboid, filopodial, mesenchymal, and collective amoeboid movements, among others (Christiansen and Rajasekaran, 2006; Friedl *et al.*, 2012;

Pagès *et al.*, 2020; Halder *et al.*, 2021). It could be no different, since embryo and cancer were created in a moment of equilibrium of the Neoplastic force. Hence why in cancer we can observe our evolutionary origin. Collective migratory movements of cancer cells are a reflection of history at the beginning of animal life. It is perfectly understandable that, from this reasoning, emerged the idea that embryonic development is recapitulated in most types of epithelial and mesenchymal cancers (Friedl *et al.*, 1995; Friedl, Hegerfeldt and Tusch, 2004; Friedl and Gilmour, 2009; Friedl and Alexander, 2011). From my viewpoint, cancers (diseases) recapitulate their own origin (Neoplasia) as an evolutionary engine.

As the epiboly proceeds to surround the ctenophore embryo, ECM assembly is initiated. Data from zebrafish research indicate that a fibronectin and laminin matrix is assembled at 65% epiboly (Latimer and Jessen, 2010). Next, i will present the details of the second exaptation within the NFM toward the second great revolution of the first embryo.

**ECM assembly and remodeling: cancer and embryogenesis are two sides of the same coin**

The ECM plays essential roles in many processes during embryonic development, such as morphogenesis, cell differentiation, migration, proliferation, and apoptosis (Adams, 2013; Walma and Yamada, 2020). These different functions of the ECM are in accord with the proposal of co-option of ECM assembly and remodeling elements in the context of Neoplasia for the formation of the first embryo. Also, as predicted in my hypothesis, ECMs play an important role in embryonic morphogenesis (Rozario and DeSimone, 2010) and cancer (Cox, 2021). I speculate that the co-option of matrix metalloproteinases (MMPs) provided the conditions for the second great revolution, which includes movements of involution, ingression, convergent extension, epithelial–mesenchymal transition (EMT), and

mesenchymal–epithelial transition (MET). I also adhere to the concept that ECM degradation or remodeling modifies not only tissue structure but also cell function and behavior (Walma and Yamada, 2020). Therefore, MMP activity needs to be highly regulated (Itoh and Nagase, 2002).

The ECM of ctenophores, as compared with that of other animals, has a different protein composition. Nonetheless, all ctenophores contain collagen IV (Fidler *et al.*, 2018), with differences in basal lamina structure between groups (Fidler *et al.*, 2017). These differences in ECM may be understood as an indication that ctenophores evolved before Porifera (Draper, Shoemark and Adams, 2019). On the other hand, studies on *Euplokamis*, *Pleurobrachia* spp., and *B. abyssicola* showed a connection between mesoglea, which contains networks of collagen fibrils, and a broad range of animal cell types, such as muscle fibers, mesenchymal cells, and nerve cells (Mackie, Mills and Singla, 1988; Jager *et al.*, 2011; Norekian and Moroz, 2019b, 2019a). A microscopy study described the basement membrane that supports ectodermal cells in *P. bachei* and *B. ovata*, but such a membrane could not be visualized in *M. leidyi* (Fidler *et al.*, 2017). Another study concluded that perlecan is absent in *M. leidyi* (Warren *et al.*, 2015). On the other hand, MMPs were detected in *P. pileus*, demonstrating that remodeling systems are found in ctenophores (Marino-Puertas, Goulas and Gomis-Rüth, 2017).

Some indications of ECM remodeling have been described in ctenophores, such as a Tolloid homolog (a type of MMP) identified in *M. leidyi* (Pang *et al.*, 2011). Tolloid, together with bone morphogenetic protein 1 (BMP-1), Tolloid-like protein 1 (TLL-1), and TLL-2, is part of a family of metalloproteinases with important functions in morphogenesis (Bayley *et al.*, 2016). BMP-1 cleaves procollagen types I–III to produce the largest fibrillar structures of vertebrate ECMs (Kessler *et al.*, 1996). Tolloid cleaves Chordin, releasing active BMPs; this is one of the crucial pathways relating ECM remodeling to cell differentiation (Marqués *et* *al.*, 1997; Streuli, 1999). Tolloid can also cleave procollagens from the ECM (Kessler *et al.*, 1996) and decorin (von Marschall and Fisher, 2010), which is known to bind to transforming growth factor (TGF)- (Noble, Harper and Border, 1992). Through these activities, Tolloid proteinases exert important functions in normal tissue assembly, embryonic patterning, and maintenance of tissue homeostasis (Troilo *et al.*, 2016). The Tolloid protein of *M. leidyi* appears to play a role in cell differentiation because it shares highly overlapping expression domains with TGF- , apparently having a similar function to that of vertebrates and invertebrates (e.g., *D. melanogaster*) (Pang *et al.*, 2011).

As could not be otherwise in the context of my hypothesis, given that embryo and cancer are two sides of the same coin, the BMP family has recently emerged as a group of proteins related to the pathogenesis of multiple types of cancers. Increased BMP-1 expression may enhance the invasiveness of gastric cancer (Hsieh *et al.*, 2018), lung cancer (Wu *et al.*, 2014), colon cancer (Sharafeldin *et al.*, 2015), and clear cell renal cell carcinoma (ccRCC). ccRCC is the renal cell carcinoma subtype with the highest rates of mortality, invasion, and metastasis (Xiao *et al.*, 2020). In ccRCC, BMP-1 knockdown suppressed malignancy both *in vitro* and *in vivo*. Furthermore, BMP-1 was detected among several other desmoplastic markers secreted by fibroblasts associated with colorectal cancer (Torres *et al.*, 2013). Finally, a strong association was found between *TLL1* gene expression and the development of hepatocellular carcinoma (Matsuura *et al.*, 2017; Mangoud *et al.*, 2021).

For an extensive review of the nomenclature of MMPs, I recommend the studies of Egeblad and Werb, 2002; Itoh and Nagase, 2002; Marino-Puertas, Goulas and Gomis-Rüth, 2017; Liu et al., 2020. MMP nomenclature can be confusing because these proteins are currently grouped according to structural characteristics but used to be classified into collagenases, gelatinases, stromelysins, and matrilysins based on their specificity for ECM components.

Despite the limited number of studies investigating MMPs in ctenophores, there are vast experimental data from other animal models and cell cultures showing a relationship between ECM remodeling and cell proliferation, thereby lending support to my proposal of an NFM. In addition to cleaving structural components of the ECM, MMPs participate in the release of cell membrane-bound precursors of growth factors, including TGF- (Peschon *et al.*, 1998). It has also been shown that IGFs sequestered by ECM proteins become bioavailable when these proteins are degraded by MMPs (Mañes *et al.*, 1997). Finally, MMPs seem to indirectly regulate proliferative signals via integrin modulation (Agrez *et al.*, 1994).

A relevant factor for my hypothesis is the ability of MMPs to facilitate cell migration, which would be key in the context of the formation of the first embryo and multicellularity. Surprisingly, ECM remodeling by MMPs seems to generate protein fragments with new functions. For instance, cleavage of laminin-5 (Giannelli *et al.*, 1997) and collagen IV (Xu *et al.*, 2001) exposes cryptic promigratory sites that promote cell motility. As could be expected, cell adhesion molecules are also substrates of MMPs. Cleavage of E-cadherin results in fragmentation of the extracellular domain and an increase in invasive behavior (Noë *et al.*, 2001). Integrin cleavage promotes cell migration (Deryugina *et al.*, 2002). The role of MMPs in apoptosis and angiogenesis is thoroughly described in a review by Egeblad and Werb (Egeblad and Werb, 2002).

It should be noted that all embryonic processes in which MMPs are involved, such as cell proliferation, migration, and differentiation, involve changes in cell phenotype and are therefore accompanied by reorganization of the actin cytoskeleton. Nowadays, there are numerous indications of a correlation between MMP activity and reorganization dynamics of the actomyosin system, suggesting mutual regulation (Bildyug, 2016). Alteration of actin cytoskeleton organization in human trabecular meshwork cells induced by direct inhibitors of actin polymerization, such as cytochalasin D and latrunculin A, resulted in activation of MMP-2 (Sanka *et al.*, 2007). A similar effect was observed in human fibroblasts (Tomasek *et al.*, 1997). MMP-2 is a pro-migratory metalloprotease (Giannelli *et al.*, 1997), and its upregulation was linked to smooth muscle cell migration and proliferation (Pauly *et al.*, 1994; Uzui *et al.*, 2000) as well as myoblast migration (El Fahime *et al.*, 2000). Differentiation processes, which i will cover extensively in the following section, seem to be dependent on reorganization of the actin cytoskeleton and MMP expression, a relationship described in cardiac muscle cells (Bax *et al.*, 2012), mesenchymal stem cells (Mannello *et al.*, 2006; Ren *et al.*, 2020), and smooth muscle cells (Yang *et al.*, 2020).

It is crucial to emphasize that the complexity of the embryo depends on its cellular rearrangement (Solnica-Krezel, 2005), and the diversity of living forms is directly associated with such rearrangements in the first embryo. Surprisingly, ECM assembly and reshaping (Latimer and Jessen, 2010) coincide with great movements of cellular rearrangement, such as invagination and convergent extension, and thus with the profound tissue reorganization that takes place within the embryo (Sherwood, 2021). There is consensus that important rearrangements in the 3D genome structure, which are decisive for cell differentiation, are determined during embryonic morphogenesis (Zheng and Xie, 2019). In all animals, without exception, genome compartmentalization and structural complexity are acquired during morphogenesis (Hug *et al.*, 2017; Kaaij *et al.*, 2018; X. Chen *et al.*, 2019; Burton *et al.*, 2020; Collombet *et al.*, 2020; Nakamura *et al.*, 2021). When we compare, for instance, *Oryzias latipes* (medaka) and *D. rerio*, it becomes clear that the 3D genome architecture is very conserved, with maintenance of the 3D structure for over 200 million years of fish evolution (Nakamura *et al.*, 2021).

Mechanical activation has been shown to induce actin polymerization, which in turn influences chromatin rearrangement (Figure 1) and promotes changes in the F-/G-actin ratio of the nucleus (Iyer *et al.*, 2012). We speculate on the possible mechanisms that could model the internal 3D structure of the embryo: (i) physical contact of two confluent epithelia that undergo mechanical tension (Pietuch *et al.*, 2013), (ii) cortical contraction altering the 3D pattern of epithelial surfaces (van Loon *et al.*, 2020), (iii) contraction of a muscle that produces tension on an epithelium and triggers its differentiation (Zhang *et al.*, 2011), (iv) epibolic cells inducing the formation of an intercellular tissue cytoskeleton using cadherins and generating a mechanical force to which the embryo responds (Xiong *et al.*, 2014), (v) tension exerted on macromeres during epiboly in ctenophores, (vi) cells responding to textural and mechanical changes produced during ECM assembly (Sunyer *et al.*, 2016), (vii) influence of ECM assembly on cell cohesion and directional persistence along mechanical gradients (Hartman *et al.*, 2017), and (viii) impact forces exerted on the whole embryo during endoderm invagination (it can be said that the embryo itself is invaginating).

In my model, multipotent cells have a 3D chromatin structure that reflects gastrulation and morphogenesis. In this way, the structural complexity of the embryo is placed within the nucleus, being one of the conceptual bases of my model (Figure 1). Finally, depending on the impact force and mechanical load, as well as on the persistence of the impact on the nucleus, cells might develop a mechanical memory (Engler *et al.*, 2006; Heo *et al.*, 2015), which translates into structural changes in the nucleus that will guide the trajectory and consistency of developmental paths in the next generation (Figure 1). This 3D structure is received by the germ line, passed to the zygote, and remodeled from an epigenetic point of view. It is understood that this ground state (Burton and Torres-Padilla, 2010; Hackett and Surani, 2014; Flyamer *et al.*, 2017; Nicetto *et al.*, 2019; Burton *et al.*, 2020) implies totipotency.

In my proposal, ECM exaptation and remodeling is considered the second great revolution because it provided the conditions for the creation of cellular arrangements that blossomed into a diversity of tissue forms, with a direct impact on cell differentiation. In zebrafish, there is evidence of differences in the amount of ECM between ventral (smaller) and dorsal (larger) regions of the embryo (Latimer and Jessen, 2010). Data for *Coturnix coturnix* show the presence of a textural gradient (rough/smooth) in fibronectin networks during gastrulation and indicate that it affects the trajectory of mesendodermal precursor cells (Loganathan *et al.*, 2012). Gradient effects of the ECM are determinant for cell migration (Chelli *et al.*, 2014; Park, Kim and Levchenko, 2018) but also for cell differentiation. Soft matrices mimicking the brain are found to be neurogenic, whereas slightly stiffer matrices mimicking muscles are myogenic, and very stiff matrices are osteogenic (Engler *et al.*, 2006).

Thus, ECM and its remodeling contribute to gastrulation, a stage wherein different germ layers are formed, embryonic axes are manifested, and the embryonic body plan begins to take shape (Schauer and Heisenberg, 2021). Mechanical cues from the ECM, such as stiffness and viscoelasticity, have a major impact on the manifestation of totipotency (Discher, Mooney and Zandstra, 2009; Connelly *et al.*, 2010; Eyckmans *et al.*, 2011; Heisenberg and Bellaïche, 2013). I must also point out that ECM remodeling is fundamental to the release of growth factors involved in cell differentiation (Marqués *et al.*, 1997; Streuli, 1999), which composes a complex dynamic that drives morphogenesis (Solnica-Krezel, 2005; Williams and Solnica-Krezel, 2017). As could be expected, it is at this moment of dynamic assembly of the embryo that 3D genome compartmentalization begins, dependent on cohesins and CTCF (Kaaij *et al.*, 2018; X. Chen *et al.*, 2019; Nakamura *et al.*, 2021). Cohesins and CTCF regulate gene expression in embryonic stem cells (Varun *et al.*, 2015; Rhodes *et al.*, 2020) and human cells (Zuin *et al.*, 2014). Cohesins are molecules dependent on mechanical strength (Kim *et al.*, 2019). This allows us to affirm, with a high degree of confidence, that it is the physical construction of the embryo that reveals its totipotency.

Cellular mechanotransduction systems have been recently shown to trigger gastrulation in *D. melanogaster*. In this animal, morphogenesis is thought to arise from coordinated changes in cell shape with participation of integrins, driven by actomyosin contractions in the embryonic mesoderm, in a Fog-dependent (Bailles *et al.*, 2019) or -independent (Mitrossilis *et al.*, 2017) manner. Therefore, the endoderm would arise from a mechanically driven cycle of cell deformation, independent of gastrulation genes such as *twist* and *snail*, which are considered to participate in morphogenetic control (Gheisari, Aakhte and Müller, 2020). This gene-independent cellular deformation is compatible with pulsatile actomyosin networks present in a wide variety of animal species (Munro, Nance and Priess, 2004; Martin, Kaschube and Wieschaus, 2009; Solon *et al.*, 2009; Rauzi, Lenne and Lecuit, 2010; Kim and Davidson, 2011; Roh-Johnson *et al.*, 2012; Maître *et al.*, 2015; Michaux *et al.*, 2018). Furthermore, in species such as *D. melanogaster* and *Caenorhabditis elegans*, Ras homolog family member A (RhoA) and myosin II (MyoII) were demonstrated to participate in gene-independent cellular deformation processes (Mitrossilis *et al.*, 2017; Michaux *et al.*, 2018; Bailles *et al.*, 2019). Rho proteins and RhoGAP domains (Rho GTPase activating proteins) have been extensively characterized in unicellular holozoans (Suga *et al.*, 2014; Sebé-Pedrós *et al.*, 2016).

Mechanotransduction seems to have, since the origin of metazoans, a relevant role as a physical mechanism responsible for morphogenesis. Cnidarians, as representatives of basal groups, are of crucial importance for understanding the evolutionary transition from diploblasts to triploblasts. Recently, the species *Nematostella vectensis* (a model organism of the phylum Cnidaria) was used to investigate the role of mechanotransduction systems in gastrulation (Pukhlyakova *et al.*, 2018). The early stages of *Nematostella* development are characterized by significant variability in cleavage patterns (Fritzenwanker *et al.*, 2007), which form an organized epithelial coeloblastula (archiblastula) (Metchnikoff, 1886). The blastula undergoes a series of extensive invaginations and evaginations correlated with cell cycle phases (Fritzenwanker *et al.*, 2007). Gastrulation occurs by a combination of invagination and late immigration involving EMT, according to an ultrastructural study (Kraus and Technau, 2006). Traditionally, gastrulation forms an endomesoderm via invagination of the blastula, which has a very expressive ECM, forming a diploblastic embryo (Fritzenwanker *et al.*, 2007).

On the other hand, the invagination process in *Nematostella* seems to reflect a highly evolutionarily conserved mechanism, with expression of the gene *brachyury* around the blastopore (Smith *et al.*, 1991; Gross and McClay, 2001; Technau, 2001; Yamada *et al.*, 2007a), whose regulation is dependent on $\beta$-catenin signaling (Arnold *et al.*, 2000; Vonica and Gumbiner, 2002). Surprisingly, blockage of MyoII-dependent gastrulation led to *brachyury* downregulation; expression of the gene could be rescued by applying external mechanical stress (Pukhlyakova *et al.*, 2018). Recovery of *brachyury* expression by an external deformation force was found to be $\beta$-catenin-dependent (Pukhlyakova *et al.*, 2018). Similarly, a study with triploblastic embryonic models demonstrated that mechanical deformation or application of a magnetic field, to mimic epiboly, was able to induce $\beta$-catenin nuclear translocation (Brunet *et al.*, 2013). Mechanical forces are also responsible for the expression of *notail* (*brachyury* homolog) in *D. rerio* and *twist* (mesodermal marker) in *D. melanogaster* (Brunet *et al.*, 2013). Overall, these studies suggest that mechanotransduction processes involved in invagination are highly conserved in metazoans.

My hypothesis of an NFM that recruited mechanosensing systems during the formation of the first embryo is consistent with the presence of a mechanotransduction system in all metazoans. On the other hand, physical forces would be participating in embryonic events that result in cellular rearrangements and that are dependent on ECM remodeling. Studies on *N. vectensis* revealed a well-consolidated ECM in the blastula (Fritzenwanker *et al.*, 2007) and an incomplete form of EMT (Kraus and Technau, 2006). I speculate that there are ECM remodeling phenomena behind mechanotransduction processes, undoubtedly indicating the second great revolution of metazoans. Expression of the *brachyury* homolog was observed in Ctenophora, in ectodermal cells surrounding the blastopore (Yamada *et al.*, 2007b), but there are still no physical studies that associate its expression with mechanotransduction or ECM remodeling mechanisms. Unfortunately, the role of MMPs in -catenin nuclear translocation via mechanotransduction has not been investigated (Brunet *et al.*, 2013; Pukhlyakova *et al.*, 2018). Nevertheless, there are numerous scientific articles relating ECM remodeling to mechanotransduction phenomena (Aitken *et al.*, 2006; Remya and Nair, 2020). Thus, external biomechanical stimuli translate into biochemical signals that initiate cellular processes such as growth, proliferation, cell differentiation (Yeh *et al.*, 2017; Yanagisawa and Yokoyama, 2021; Sthanam *et al.*, 2022), and embryonic morphogenesis (Brunet *et al.*, 2013; Mitrossilis *et al.*, 2017; Pukhlyakova *et al.*, 2018; Bailles *et al.*, 2019).

My hypothesis proposes that cancer and embryo were established together during evolution, and i can predict that *brachyury* expression is linked to cancer events. In cancer biology, *brachyury* has been associated with EMT, resulting in a more invasive tumor phenotype (Fernando *et al.*, 2010). The idea that *brachyury* is the driver of EMT is well-established (Chen *et al.*, 2020) for various types of carcinoma (Miettinen *et al.*, 2015). In fact, the protein has been explored as a target antigen in tumor vaccines (Palena *et al.*, 2007). Brachyury expression is often associated with more aggressive forms of cancer and poor prognosis. Consistent with its embryonic location in the notochord (Schulte-Merker *et al.*, 1992; Hotta *et al.*, 2000), *brachyury* has been suggested to be crucial for chordoma development and spread (Walcott *et al.*, 2012). In my view, Brachyury is part of a very ancient framework involved in the formation of the first embryo. Its location in the blastopore of Ctenophora (Yamada *et al.*, 2010) and role in the mechanotransduction system of Cnidaria (Pukhlyakova *et al.*, 2018) are factors that determine its early link with cancer.

As expected, currently, there seems to be a consensus that mechanical stress is associated with morphogenesis and cancer (Paszek and Weaver, 2004). Developing structures are subject to a multitude of tensile forces that shape morphology and differentiation, as will be addressed in the following section (Ingber, 2006; Wozniak and Chen, 2009; Farge, 2011; Anandasivam, 2020; Gillard and Röper, 2020; Kindberg, Hu and Bush, 2020; Seo *et al.*, 2020; Tsata and Beis, 2020; Villedieu, Bosveld and Bellaïche, 2020; Kim *et al.*, 2021; Narayanan, Mendieta-Serrano and Saunders, 2021). Malignant transformation is also associated with dramatic changes in tension, which include elevated compressive forces, mechanoreciprocity, and increased ECM stiffness (Paszek and Weaver, 2004; Paszek *et al.*, 2005; Aguilar-Cuenca, Juanes-García and Vicente-Manzanares, 2014; Fattet *et al.*, 2020; Reuten *et al.*, 2021). The chronic increase of these tension forces influences tumor growth, tissue morphogenesis, and invasion. Matthew Paszek, in describing the impact of mechanical stress extension and persistence on mammary gland morphogenesis, stated that õa chronic increase in cytoskeletal tension, mediated by sustained matrix stiffnessí if of sufficient magnitude and duration, could drive the assembly/stabilization of focal adhesions to enhance growth and perturb tissue organization, thereby promoting malignant transformation of a tissueö (Paszek *et al.*, 2005). In the context of my model, stress of sufficient magnitude and duration contributed to EMT and morphogenesis of the first embryo. Other interesting articles showed that increased ECM stiffness (caused by collagen crosslinking) induced invasion of premalignant mammary epithelium (Levental *et al.*, 2009; Wei *et al.*, 2015) as well as EMT in oral squamous cell carcinoma (Matte *et al.*, 2018).

Therefore, the connection of EMT with cancer (Thiery, 2002; Seiki, 2003; Christiansen and Rajasekaran, 2006; Thiery and Sleeman, 2006; Esquer *et al.*, 2021) and embryology (Morrissey and Sherwood, 2015; Sherwood, 2021) is indisputable and well-established in the scientific literature, being, without a shadow of a doubt, one of the main reasons why developmental biologists venture into cancer research. The supposed contradictions of classical EMT as a universal resource in tumor invasion and metastasis, which include incomplete EMT, reversion to epithelial phenotype, and collective migration, may be elucidated in the field of embryology. Incomplete EMT occurs in embryonic development of *N. vectensis* (Kraus and Technau, 2006), reversion to an epithelial phenotype is observed in MET for the formation of various embryonic structures (Chaffer, Thompson and Williams, 2007; Tam *et al.*, 2007; Pitsidianaki *et al.*, 2021), and, finally, collective migration finds support in embryonic morphogenetic movements (Rørth, 2007; Amy *et al.*, 2008; Weijer, 2009; Bloomekatz *et al.*, 2012; Omelchenko, Hall and Anderson, 2020).

It is also important to discuss the regulatory systems of MMPs. Research on MMPs and EMT is very controversial mainly because of the complexity and various roles of MMPs (Gialeli, Theocharis and Karamanos, 2011). Therapeutic strategies using MMP blockers have not been successful in the field of oncology. MMPs may also be regulated by inhibitors, some of which promote invasive activity, whereas others seem to suppress such activity (Egeblad and Werb, 2002). These factors reinforce the idea that MMPs play a role in cellular homeostasis and ECM assembly (Bayley *et al.*, 2016). Some Tolloid-like metalloproteinases are important for normal tissue formation (Troilo *et al.*, 2016). Therefore, the idea of ECM assembly and remodeling is supported by embryology and is one of the crucial aspects of my NFM proposal.

To conclude my reflections on ECM, it is important to consider the amoeboid traits of unicellular holozoans (Mendoza, Taylor and Ajello, 2002; Paps and Ruiz-Trillo, 2010; Sebé-Pedrós, Irimia, *et al.*, 2013; Suga and Ruiz-Trillo, 2015; Hehenberger *et al.*, 2017). My hypothesis suggests that fertilization (anisogamy), NFM recruitment (involving adhesion), and ECM assembly and remodeling are part of a biological switch capable of repressing amoeboid characteristics within the embryo construction process. It is not surprising that ECM remodeling allows revealing amoeboid traits in an embryonic context, given the phylogenetic origin of the first embryo. The three main cell types that have amoeboid movements are primordial germ cells (which receive the neoplastic module) (Jiang, Clark and Renfree, 1997; Mazzoni and Quagio-Grassiotto, 2021), immune system cells (Norberg, 1970), and stem cells (Rinkevich *et al.*, 2022). In cancer research, it was demonstrated for the first time that amoeboid movement is not a means of migrating but rather a cellular state (Graziani *et al.*, 2022; Mohammadalipour *et al.*, 2022). Currently, mesenchymal–amoeboid (Taddei *et al.*, 2014), epithelial–amoeboid (Crosas-Molist *et al.*, 2017), and epithelial–mesenchymal–amoeboid transitions (Emad *et al.*, 2020) are recognized in some cancers. These transitions are consistent with my hypothesis about the evolutionary origin of unicellular Holozoa. Epithelial–mesenchymal–amoeboid last transitions bring to mind embryo cells that carry (Czirok *et al.*, 2006; Davidson *et al.*, 2008; Zamir, Rongish and Little, 2008) and modify (Mazzoni and Quagio-Grassiotto, 2021) the ECM while migrating.

In the following section, i will address what is perhaps one of the most obvious aspects of NFM: cell differentiation. Embryogenesis reveals and differentiates the distinct types of animal cells (Gilbert, 2017), and cancer is traditionally known as a disease of cell differentiation (Markert, 1968).

**Differentiation in embryo and cancer**

Implicitly, i am proposing that morphogenetic changes occur simultaneously with cell differentiation during embryogenesis. Therefore, form of organization and embryonic construction have a direct impact on differentiation. Studies have shown that the nanostructured morphology of the cell surface is a key factor that promotes neurogenesis; for instance, highly spiked structures are more efficient in inducing neuronal differentiation (Poudineh *et al.*, 2018). The geometry and dimension of the substrate (topography) also seem to guide neuronal or glial differentiation (Ankam *et al.*, 2013). It has also been shown that cell shape, cytoskeletal tension, and RhoA signaling proteins are determinants of human mesenchymal stem cell (hMSC) differentiation into adipocytes or osteoblasts (McBeath *et al.*, 2004). A surprising result was that actin cytoskeleton remodeling transduces mechanical stimuli that promote muscle cell differentiation involving RhoA/ROCK signaling (Huang *et al.*, 2012). hMSCs with functional blockade of vinculin do not express MyoD, but RUNX2 expression (involved in osteoblast differentiation) is stimulated, indicating that an adhesion protein sensitive to mechanical force can regulate the fate of stem cells (Holle *et al.*, 2013).

Understanding the factors, including mechanical ones, that direct germ layer organization during development is one of the main goals of developmental biology. Michael Krieg, in a study using atomic force microscopy to quantify adhesive and mechanical properties of the ectoderm, mesoderm, and endoderm of progenitor cells of zebrafish embryos, showed that the differential tension of the actomyosin-dependent cellular cortex, regulated by Nodal/TGF- signaling, constitutes a key factor in germ layer organization during gastrulation (Krieg *et al.*, 2008). In the context of mechanical tension, Adam Engler and Dennis Discher demonstrated that mesenchymal stem cells, influenced by ECM elasticity and MyoII, commit to neural, muscular, or osteogenic lineages (Engler *et al.*, 2006). In my view, perception of elasticity is a mechanical tension phenomenon that results in mechanotransduction, which is directly involved in cell differentiation. As previously mentioned, soft matrices that mimic the brain are neurogenic, slightly stiffer matrices that mimic the muscle are myogenic, and very stiff matrices are osteogenic (Engler *et al.*, 2006).

Thus, i am suggesting that the mechanical environment might have been a determinant factor for regulation of cell differentiation in the first embryo. In other words, cells can perceive differences in embryonic microenvironments and respond by differentiating into different cell types and forms of tissue organization (Wang, Butler and Ingber, 1993; Gumbiner, 1996, 2005). Some physical and mechanical cues include cell form (Poudineh *et* *al.*, 2018), cytoskeleton tension (McBeath *et al.*, 2004), mechanotransduction pathways involving RhoA or YAP/TAZ (McBeath *et al.*, 2004; Huang *et al.*, 2012; Piccolo, Dupont and Cordenonsi, 2014), ECM stiffness (Dupont *et al.*, 2011), ECM composition (Hartman *et al.*, 2017), and ECM topography (Ankam *et al.*, 2013), which might mimic the path of migrating cells.

I speculate that some mechanical clues can be found in the morphology of the first embryo. The great embryological revolution brought by epiboly must have produced, because of its integrated and collective movements, mechanical tension (mechanotransduction) on the actin cytoskeleton of macromeres. Such tension is produced in the opposite direction of epiboly, in the aboral direction (Figure 2). In a model of collective cancer cell migration, a viscoelastic resistive force was detected and found to always be directed opposite to the migration direction (Pajic-Lijakovic and Milivojevic, 2020). A mechanical wave would be generated in the direction of epiboly-stimulated migration (Pajic-Lijakovic and Milivojevic, 2022), possibly having implications for embryo patterning (Serra-Picamal *et al.*, 2012). I suppose that the tension on macromeres would support the integrated movement of micromeres (epiboly). In cancer models, response mechanical tensions are clearly defined as reciprocal stress forces or mechanoreciprocity (Butcher, Alliston and Weaver, 2009). I must also take into account that the ECM is assembled as epiboly progresses (Latimer and Jessen, 2010) and that there is evidence of a gradient of ECM stiffness in embryogenesis (Loganathan *et al.*, 2012). This might have been determinant to direct the migratory movement of cells in epiboly (collective cell durotaxis) (Sunyer *et al.*, 2016) in a context of long-range force transmission (Serra-Picamal *et al.*, 2012). Thus, it is assumed that mechanical alterations in migratory cells and macromeres could be responsible for the initial embryonic patterning in the absence of Wnt or other morphogens.

It should be noted that when cells undergoing epiboly reach the oral region, macromeres are subject to numerous tensions of different magnitudes, frequencies, and durations (Figure 2). As demonstrated by using cancer models, these tension forces, depending on their magnitude and persistence, are capable of improving cell growth, disrupting tissue organization, and producing EMT (Paszek *et al.*, 2005). In this context, I propose that migratory cells undergoing epiboly create a mechanical force that could be directly linked to endoderm invagination (Figure 2). Consistent with the idea of tension exerted on macromeres, Mark Martindale detected an embryo curvature that gives rise to a concave disc with an opening at the oral pole. It is unknown whether the force required to produce this change in shape is derived from macromeres or underlying micromeres (Martindale and Henry, 2015). In my view, this curvature is a harbinger of the effects of mechanotransduction (mechanoreciprocity) on macromeres, which will subsequently divide (forming oral micromeres), invaginate (forming the endoderm), and contribute to the invagination of ectodermal cells (forming the pharynx and esophagus).

Other mechanical clues of cell differentiation include contact between tissues and how they transmit and transform mechanical tension into chemical signals. Studies on tension-induced mechanotransduction in *C. elegans* demonstrated the influence of muscle cells on epithelial morphogenesis (Zhang *et al.*, 2011). In the referred study, contact between muscle and epithelium was proposed to occur through hemidesmosomes, serving not only as fixation structures but also as mechanosensors that respond to tension, thereby triggering signaling processes that involve Rac GTPase, p21-activated kinase (PAK-1), the adapter Git-1, and PIX-1 (Zhang *et al.*, 2011). In the context of my hypothesis, in ctenophores, there are no data corroborating the influence of muscle cells on comb row formation, but a connection of gonadal tissues with comb row formation has been suggested. Gametogenic tissues occur in the eight meridional canals underlying comb rows or in their homologs or derivatives (Pianka, 1974). Endodermal formation of the gonad is well established, and gonadal organization allows distinguishing the different types of ctenophores (Pianka, 1974).

Tissue contact and interaction, as occurs in embryonic induction, are among the most important processes in modern embryology (Spemann and Mangold, 1924; Gurdon, 1987; Jessell and Melton, 1992; Slack, 1993; Hemmati-Brivanlou and Melton, 1997; Harland, 2000). It is not surprising that an important part of morphogens are linked to the ECM and its remodeling. The only aspects that differ from mechanotransduction are the approach to the living tissue constitution and interactions of an embryo with conserved integrity. Thus, in view of mechanotransduction, it is possible to find several clues of differentiation in the contacts made by tissues during morphogenesis. For instance, micromere cells that move by epiboly differentiate into epidermal cells. Macromere cells that undergo stress form endodermal cells. Comb row formation requires inductive signals between $m_1$ and $e_1$ lineages (Martindale and Henry, 1997). Classical inductive signals from E and M macromeres were proposed by Henry and Martindale (2001). Oral micromeres, on the other hand, migrate and undergo MET to form muscle cells. Finally, if oral micromeres come into contact with the aboral epidermal surface, they form lithocytes of the apical organ (Martindale and Henry, 1999). Classical inductive signals are well described in ctenophores (Farfaglio, 1963; Martindale, 1986; Martindale and Henry, 1996, 1997), as are "embryonic fields or equivalence groups, controlled by cell–cell interactions" (Henry and Martindale, 2004).

In ctenophores, gastrulation produces the ectoderm by epiboly of aboral micromeres and the endoderm by invagination (embolism) of macromeres that turn inward, carrying some micromeres into the inside of the embryo. These micromeres, known as gastrular (Farfaglio, 1963) or oral (Martindale and Henry, 2015) micromeres, are produced by macromere division (Pianka, 1974). They appear on the oral side and are brought inside the embryo by subsequent macromere invagination. Studies on *M. leidyi* showed that most oral micromeres are progenitors of apical organ lithocytes (Freeman and Reynolds, 1973) and give rise to mesodermal musculature (Martindale and Henry, 1999). It should be noted that oral micromeres can also contribute to endodermal derivatives, and this cannot be ruled out by technical limitations (Martindale and Henry, 1999). The architecture and organization of muscle cells in *M. leidyi* is singular in that it follows an aboral–oral orientation (longitudinal muscles, see Figure 7F in Martindale and Henry, 1999), in the direction that the epidermis was constructed. Therefore, i speculate that mesenchymal cells harbor the niches of epidermis differentiation or even the niches of nervous system differentiation.

Another mechanical cue of differentiation is ECM rigidity. The presence of an ECM textural gradient in gastrulating embryos may affect cell trajectories (Loganathan *et al.*, 2012). Qualitative differences in texture (rough/smooth) on a spatial scale can translate into cellular differentiation. In the context of mechanical stress, mesenchymal stem cells commit to neural, muscular, or osteogenic lineages depending directly on the elasticity of the ECM and MyoII (Engler *et al.*, 2006). Consistent with Adam Engler's studies, in ctenophores, oral micromeres differentiate into neural cells (Jager *et al.*, 2011), muscle cells (Martindale and Henry, 1999), and mesenchymal cells of the mesoglea (Pang and Martindale, 2008) after a migratory process. In *P. pileus*, two distinct nerve networks are observed: a mesogleal nerve network, loosely organized throughout the mesoglea of the body, and a much more compact "nerve" network with polygonal meshes in the ectodermal epithelium (Jager *et al.*, 2011). It is well known that, in both vertebrates (Mongera *et al.*, 2019; Sambasivan and Steventon, 2021; Wymeersch, Wilson and Tsakiridis, 2021) and ascidians (Hudson and Yasuo, 2021), some posterior neural tissues share a common origin with the mesoderm, although the mechanisms of specification are different (Hudson and Yasuo, 2021). It seems logical to think, on the basis of the evaluated data, that a neuromesodermal lineage has been conserved since the beginning of metazoans. The spatiotemporal localization of the neuromesodermal lineage in vertebrates generally coincides with co-expression of the transcription factors brachyury/TbxT and Sox2 (Aires, Dias and Mallo, 2018). Brachyury was characterized in ctenophores and cnidarians (Yamada *et al.*, 2007b, 2010; Pukhlyakova *et al.*, 2018). Sox2 was identified in ctenophores, associated with regions of intense cell proliferation (Schnitzler *et al.*, 2014; Moroz, 2015), neurosensory epithelia (Jager *et al.*, 2008), and cancer (Dey *et al.*, 2022; Pouremamali *et al.*, 2022). Thus, i speculate that systems that induce the nervous system are conserved in invertebrates (Gilbert, 2000), and that epiboly triggers the formation of a preneural territory by default.

Nervous system formation is one of the most enigmatic points in the embryology of ctenophores; its independent origin has generated much controversy (Marlow and Arendt, 2014; Moroz *et al.*, 2014). In my hypothesis, two great revolutions took part in embryo formation, namely epiboly and the ECM. These revolutions serve as an inspiration for our understanding of nervous system formation. The cellular bases of the default theory for neural induction are well established (Stern, 2006; Levine and Brivanlou, 2007). It is natural to search for the genes responsible for the process or to try to understand, at least, whether there is any relationship between these genes and nervous system formation in primitive animals such as ctenophores. Surprisingly, almost all components are present in early embryonic development. Tolloid and TGF-β appear in overlapping patterns, consistent with epidermal induction preferentially in the aboral region (identified by Martindale as the tentacular bulb) (Pang *et al.*, 2011). BMP homologs also occur in the aboral region and are related to the apical organ (Pang *et al.*, 2011). The nervous system appears with a distribution typical of bilaterians, with an aboral neurosensory complex (Jager *et al.*, 2011). The problem is that BMP inhibitor genes such as *chordin* and *follistatin* (Pang *et al.*, 2011) and BMP and TGF-β inhibitor genes known to act in the endoderm region (Henry *et al.*, 1996) are missing.

Contributing to the deconstruction of the idea of gene participation, Juan Larraín described that two cysteine-rich domains, rather than the entire chordin protein, are responsible for inhibiting BMP (Larrain *et al.*, 2000). Cysteine-rich domains in other intracellular proteins have been shown to be one of the great innovations of unicellular holozoans (Herman *et al.*, 2018). Proteins with these domains, such as integrins, undergo conformational changes to reveal cryptic sites not predicted in the initial structure; therefore, the dynamics in which these proteins interact seem to be very important (Beglova *et al.*, 2002). Proteolytic processing of some proteins implicated in embryonic development, however, may reveal new biochemical activities associated with BMP inhibition (Yu *et al.*, 2000). In this context, the domains of interaction with BMP are important in the ECM (Sedlmeier and Sleeman, 2017). Several components of the ECM bind to BMP. Type IIA procollagen containing cysteine-rich propeptide binds to TGF- 1 and BMP-2 (Zhu *et al.*, 1999) and collagen IV binds to BMP-2 (Paralkar *et al.*, 1992) and the BMP-2/4 homolog decapentaplegic in *D. melanogaster* (Wang *et al.*, 2008). The physical cues of BMP gradients and their relationship with mechanotransduction are found in the ECM (Sedlmeier and Sleeman, 2017). BMP diffusion coefficients are reduced from about 87 to 0.10 $\mu m^2/s$ in the presence of collagen IV (Sedlmeier and Sleeman, 2017). Integrins bind to the BMP receptor and form a mechanoreceptor complex that mediates Smad1/5/8 phosphorylation in response to compressive forces (Zhou *et al.*, 2013). BMP-1 receptor interacts with integrin subunit 1 to induce Smad1/5 signaling in response to increased substrate stiffness (Guo *et al.*, 2016), thus becoming a mechanosensory system.

At the embryonic level, ECM networks provide highly dynamic material properties necessary to accommodate the large-scale deformations and forces that shape embryos. In models of epithelial cancer cells, tension waves are generated and extend over long distances (Serra-Picamal *et al.*, 2012) in the direction of migration, functioning as "morphogens" (Morita *et al.*, 2017; Das *et al.*, 2019; Agarwal and Zaidel-Bar, 2021). In ctenophores, the nervous system is located predominantly in the aboral region, indicative of the effects of early patterning (Jager *et al.*, 2011); however, this was not always so. *Ctenorhabdotus campanelliformis* had longitudinal axons connecting the apical organ and ciliated sulci (aboral) to a circumoral (oral) nerve ring (Parry *et al.*, 2021). Also, the ECM microenvironment changes markedly in time and space during morphogenesis, producing fluctuations in textural properties (Loganathan *et al.*, 2012). These textural gradients stem from physical forces involved in the self-organization of the first embryo. Thus, i speculate that the ECM sequesters BMP according to textural patterns, with regions of BMP diffusion that induce the epidermis and regions without BMP diffusion (sequestered into the ECM) that allow default formation of the nervous system. The profile of polygonal networks of the nervous system would follow the textural pattern of the ECM in the absence of BMP-4 diffusion (Jager *et al.*, 2011). This same textural pattern might have been important for induction of a neuromesodermal lineage during gastrulation.

Unfortunately, embryological models of pattern formation disregard mechanical morphogenesis (Turing, 1952) and focus exclusively on chemical morphogenesis (Gierer and Meinhardt, 1972), not being used to identify explanatory mechanisms of self-organization and asymmetries in the first animal embryo (Wolpert, 1969, 1971). Nevertheless, i identified points of convergence between chemical models and mechanical morphogenesis. Alan Turing believed that morphogens needed to specify only two states, a new state and a default state (Turing, 1952). Chemical models consider short-range activators and long-range inhibitors (Turing, 1952; Gierer and Meinhardt, 1972). Mechanical phenomena produce long-range tensile forces (Serra-Picamal *et al.*, 2012) and ECM compositional effects that change BMP diffusion coefficients (Sedlmeier and Sleeman, 2017), resembling short-range activators. Furthermore, small random molecular fluctuations can slightly increase the concentration of activators at certain positions (Gierer and Meinhardt, 1972). Mechanical waves and textural fluctuations fit perfectly into this expectation of a chemical model of morphogens (Turing, 1952). I believe that a physical approach to the embryonic development of ctenophores will bring, in the near future, important answers for the field of biology.

Let us reflect on Alan Turing's work on the chemical bases of morphogenesis, in which he "proposed to give attention rather to cases where the mechanical aspect can be ignored and the chemical aspect is the most significant" (Turing, 1952). It is very difficult to find embryological (Zhang *et al.*, 2011; Tlili *et al.*, 2019) or cancer (Hirway, Lemmon and Weinberg, 2021) phenomena where mechanical aspects can be ignored, but, at the time of Turing's ideas, the focus on the search for genes to understand morphogenesis was justified by the challenges and limitations of that period (genes = information). Today, i have a growing theoretical basis of the physics of embryonic development, which helps to understand that physical forces do not evolve; they are universal and can change the biological components that perform a certain force, but the end result is always the same. Chordin and collagen IV regulating BMPs have the same effect on nervous system induction. Physical phenomena are modulable, specific in the context of interactions that arise within an embryo, modifiable by the environment, and absolutely determinant of animal evolution because of their preponderant participation in the construction of an NFM.

Thus, returning to my hypothesis, macromeres subjected to a tensile force during gastrulation will differentiate into an endodermal layer and all mesodermal constituents of the mesoglea that include a neural network. In this way, my goal is to indicate that it is embryogenesis that builds tissue architecture and organization. As long as we do not consider the embryonic process, we will continue to affirm that totipotency is the ability to generate distinct cell types, to the detriment of thinking of totipotency as the result of the embryo generating tissue architecture and organization during morphogenesis. In this reasoning, epiboly, invagination, and all morphogenetic movements would be directly involved in revealing totipotency.

Defining totipotency as exclusively a differentiation problem without considering morphogenesis seems to be the result of the experimental approach to the problem. We most often intend to study cell differentiation by destroying the embryo and its organization and losing the physical and mechanical relationships of the process (Discher, Mooney and Zandstra, 2009). A good example of the errors of this approach is the idea that pluripotent embryonic stem cells are unique tools for studying mammalian development and differentiation in a Petri dish (Rossant, 2008) or the possibility of establishing new models of mammalian development from isolated embryonic stem cells (Murry and Keller, 2008; Young, 2011). The ultimate aspiration is to control the potency of embryonic stem cell differentiation and direct the development of these cells along specific pathways, thereby losing relevant information contained within these embryonic cells, their niches, relationships with the ECM, and morphogenetic movements that could provide a solution to the potency problem.

Another very common error is to attribute the morphogenetic construction of the embryo to genes (Gehring, 1996). In vertebrates, gastrulation genes are part of genetic strategies of developmental biology; however, when gastrulation occurs without these genes, induced by physical forces (which we normally do not regard as entities that execute and are responsible for morphogenesis) (Brunet *et al.*, 2013; Mitrossilis *et al.*, 2017; Pukhlyakova *et al.*, 2018; Bailles *et al.*, 2019), we begin to question our technical and theoretical limitations. We also attribute to genes the establishment of differentiation power, as best exemplified by induced pluripotent stem cells (iPSCs) (Takahashi *et al.*, 2007). However, although the benefits of basic and applied studies on iPSCs are undeniable, the disadvantages of this model in its current state and, in particular, the aspects of differentiation protocols that require further refinement are commonly ignored (Ratajczak, Bujko and Wojakowski, 2016; Antonov and Novosadova, 2021). The relationships between reprogrammed embryonic stem cells and cancer are also poorly understood (Hwang, Kim and Kim, 2009; Lin *et al.*, 2014; Du *et al.*, 2020; Seno *et al.*, 2022).

In placing a Neoplasia model at the heart of evolution and embryogenesis, one of the main pillars of which is cell differentiation (Markert, 1968; Pierce, 1974), it becomes clear that iPSC formation deserves attention within my theoretical framework. Introduction of *Oct3/4*, *Sox2*, *Klf4*, and *Myc* to adult human fibroblasts led to cell reprogramming (Takahashi *et al.*, 2007) as well as activation of the NFM, which is present in all adult human cells, albeit in a silent state. *Oct3/4* (de Jong and Looijenga, 2006), *Sox2* (Dey *et al.*, 2022), *Klf4* (Qi *et al.*, 2019), and *Myc* are linked to cancer. Coherently, iPSCs injected intraperitoneally produced a teratoma, demonstrating pluripotent characteristics (NANOG expression) compatible with the denomination of embryoids (Abad *et al.*, 2013). According to my hypothesis, the expected result would be the formation of an embryo (teratoma) in the Neoplastic context of this *in vitro* system. Such an occurrence should not come as a surprise, given the somatic origin of some teratocarcinomas (Mintz, Cronmiller and Custer, 1978) and the presence of germline genes in various types of tumors (Simpson *et al.*, 2005). Of note, in an embryonic model, differentiated human cells showed different potentials to form cancer (Cofre and Abdelhay, 2017).

There is a theoretical approach to understanding iPSCs from an embryological and oncological perspective (Trosko, 2014). In his beautifully crafted essay, James Trosko (Trosko, 2014) proposed three questions: First, "can an uncontrolled proliferation of the earliest form of a single cell organism, in a microenvironment of adequate nutrients, temperature, etc., be likened to the uncontrolled growth of a cancer cell in a living human being?ö Yes, considering the close phylogenetic connection with unicellular holozoans and the molecular and developmental pathways recruited (co-opted) from unicellular holozoans. Second, "is there any evolutionary link, and if so, what might that link be?" Neoplasia is at the heart of animal evolution, and embryos and cancer develop in a period of stabilizing neoplastic strength. Third, "is cancer an evolutionary throw-back to the properties of the single cell organism?" Not directly to the unicellular organism but to the process of formation of the first embryo, as a consequence of the fusion of cells of unicellular holozoans, which resulted in co-option of an NFM.

Finally, it is necessary to reflect on the meaning of cancer as the main core of animal embryogenesis. Insofar as cellular elements were co-opted during the natural experience of embryo formation, they did so in a parsimonious way but always in a context of controlled Neoplasia. Sustained, uncontrolled growth would not make sense for embryo integrity, metabolic viability, or harmony with environmental conditions. Benign tumors of variable sizes would not be an interesting innovation in animal evolution. Therefore, a structure of ECM and cadherins would constitute a first phase of growth control (Kemler, 1993; Takeichi, 1995; Tepass *et al.*, 2000; Jeanes, Gottardi and Yap, 2008; Hegazy *et al.*, 2022). Similarly, a benign tumor that dissociates via amoeboid movements, mimicking metastasis, would not be of interest for maintaining multicellular integrity in what we call an embryo. Adhesome systems also contributed to controlling the intrinsic properties of unicellular organization in holozoans, allowing migratory amoeboid movements to occur only at specific moments of embryogenesis.

For this, a key point of animal organization was the co-option of Hippo pathways and effectors (Pan, 2007, 2010; Zhao *et al.*, 2007, 2010; Varelas *et al.*, 2010; Kim *et al.*, 2011) to mechanotransduction systems and cell differentiation (Fletcher *et al.*, 2018). This would provide the possibility of controlling proliferation and cell morphogenesis. YAP/TAZ activation is an indication of the cell's social behavior, including cell adhesion and mechanical signals received from tissue architecture and the surrounding ECM (Piccolo, Dupont and Cordenonsi, 2014). RhoA/ROCK competes with YAP "to regulate amoeboid breast cancer cell migration in response to biomechanical force" (Mohammadalipour *et al.*, 2022). Thus, YAP/TAZ appears as a centerpiece of a signaling nexus by which cells take control of their behavior according to their shape, spatial location (Piccolo, Dupont and Cordenonsi, 2014), and, in my view, embryonic context.

The fact that a direct link appears between differentiation and embryonic morphogenesis coherently indicates that germline formation could only occur during or after morphogenesis, with characteristics of multipotent stem cell populations (Juliano, Swartz and Wessel, 2010; Seervai and Wessel, 2013; Fierro-Constaín *et al.*, 2017). The embryo as a whole receives directly or indirectly physical and tension impacts; therefore, the physical, mechanical (tension actin networks), and even electrical (nerve impulses) organization of the embryo can be important in modeling the 3D structure of the nucleus of multipotent cells. The mosaic presented by ctenophores at the beginning of their development (Pianka, 1974) would represent, in my opinion, the structural polarity of the cytoskeleton of microtubules and actin, allowing asymmetric distribution of organelles and proteins; it may also be a direct reflection of cell division, which would produce unequal segregation of ectoplasm and endoplasm. In any case, I emphasize in my hypothesis that cell differentiation is a component linked to morphogenesis (Cofre and Abdelhay, 2017; Cofre, Saalfeld and Abdelhay, 2019).

At this point, it is important to clarify that mechanotransduction cannot be recruited within the structure of chromatin and the Neoplastic module. Mechanotransduction shapes the 3D structure of chromatin, recruiting components that participate in cellular systems and the mechanosensory architecture of the cell, among other cellular processes, consequently establishing an NFM and incorporating Neoplasia as a construction force of the embryo. This explains why it is very difficult to find components of developmental pathways, biomechanical force transmission, and embryonic cell organization that are not related to cancer. For example, adhesome, its interconnected components, such as YAP/TAZ (Mohammadalipour *et al.*, 2022), -catenin, Kaiso/p120, IGF1R/RhoA/ROCK (Liu *et al.*, 2022), and emerin, as well as nuclear lamina components (Liddane and Holaska, 2021), are all related to cancer. There is a finite chain of elements that act in a modular way and intertwine Neoplasia within animal organization. Therefore, in the context of my hypothesis, all components that are part of the cellular framework of Neoplasia have become essential elements in embryo constitution and animal evolution. A phylostratigraphic analysis demonstrated a link between cancer and multicellularity (Domazet-Lo-o and Tautz, 2010) but was not able to unravel the implications of this observation within the field of embryology.

According to my hypothesis, i can predict that there have been moments of frank Neoplasia in evolutionary history. For example, among metazoans, sponges and cnidarians (anemones and corals) express the largest and most incredible number of antitumor compounds (Casás-Selves and DeGregori, 2011; Rocha *et al.*, 2011; Sima and Vetvicka, 2011; Pejin *et al.*, 2013; Hu *et al.*, 2015; Oliveira *et al.*, 2018; Ilhan and Pulat, 2020). The question is what do these animals produce antitumor compounds for? Is it a simple biological curiosity, coincidence, or a response to the context that generated these animals? Probably, the most coherent answer, if my hypothesis is accepted, is that these living representatives arose from animals that experienced an intense stage of Neoplasia and morphological creativity. There is no evidence of antitumor agents in ctenophores, given the lack of studies on the topic.

Other records of intense Neoplasia are found in mitochondrial genomic recombination studies. Mitochondrial genomes show an extremely rapid rate of evolution in ctenophores, possibly constituting a peculiarity of this group of animals (Pett *et al.*, 2011; Kohn *et al.*, 2012; Arafat *et al.*, 2018). The results showed that this high evolutionary rate affects not only nucleotide substitution but also gene rearrangements that are highly saturated (Arafat *et al.*, 2018). Such a peculiarity is also found in cancer. A high number of disruptive mutations have been reported for protein and tRNA coding regions of the mitochondrial genome in human patients with Hurthle cell thyroid carcinoma (Ganly *et al.*, 2018; Gopal *et al.*, 2018). In 64 tumor samples from 55 prostate cancer patients, the mitochondrial genome showed a high mutation rate compared with autosomal chromosomes (Lindberg *et al.*, 2013). The role of mitochondrial genome mutations in cancer is far from being elucidated by current models (see review by Zong, Rabinowitz and White, 2016; Nissanka, Minczuk and Moraes, 2019). We will be able to see our evolutionary past by looking at cancer.

Another peculiarity of the mitochondrial genome, observed in *M. leidyi*, was the loss of all tRNA genes and of mitochondrial aminoacyl-tRNA synthetase encoded by the nucleus (Pett *et al.*, 2011). Surprisingly, the *atp6* gene was relocated to the nuclear genome and acquired introns and a mitochondrial targeting presequence (Pett *et al.*, 2011). As expected, this same process of somatic mitochondrial transfer is observed in cancer. A study of 559 primary cancers (including breast cancer, osteosarcoma, and prostate cancer) and 28 cancer cell lines revealed the somatic transfer of mtDNA to the nuclear genome (Ju *et al.*, 2015). Two previous studies demonstrated the integration of mtDNA fragments into Myc in HeLa cells (Shay *et al.*, 1991) as well as mtDNA transfer in germline cells (Turner *et al.*, 2003).

Although there are still no answers for questions regarding the mechanisms underlying the enormous variation of mitochondrial genes in ctenophores, I suggest that ctenophores, in their evolutionary origin, were at the extreme limit of Neoplastic pressure, as can be deduced from mitochondrial analyses of all living representatives. Furthermore, despite the limitations of obtaining complete mitochondrial genomes for systematic and evolutionary studies, an accelerated rate of evolution has been found in some groups of sponges (orders

Dendroceratida, Dictyoceratida, and Verticillitida) (Wang and Lavrov, 2008; Lavrov, 2014) and in cnidarians (eight species of octocorals) (Muthye *et al.*, 2022).

Other lines of evidence of the intense Neoplasia may be found in fossil records. Ctenophores were not always as we know them today. Evidence shows that the number of comb rows was greater than eight, as is currently well characterized in developmental biology. The oldest ctenophore fossil, called *Eoandromeda octobrachiata*, is from the Ediacaran Period. The species had eight comb rows in spiral orientation and no tentacles (Tang et al., 2011; for a different view, see Zhao et al., 2019). In the Lower Cambrian, the oldest ctenophore had eight comb rows (Chen *et al.*, 2007); for species in the Middle Cambrian, Conway Morris reported up to 80 comb rows and a body measuring 75 mm in length and 48 mm in diameter (Morris and Collins, 1996). Two species from the Middle Cambrian were identified in Utah, USA, namely *Thalassostaphylos elegans* (with 16 comb rows, an oral skirt, and an apical organ with polar fields) and *Ctenorhabdotus campanelliformis* (with 24 comb rows, an oral skirt, an apical organ enclosed by a capsule, and very well preserved neurological tissues), demonstrating that Cambrian ctenophores had a more complex neuroanatomy (Parry *et al.*, 2021) than extant species (Jager *et al.*, 2011; Norekian and Moroz, 2019b, 2019a). Thus, NP manifested itself in bodily innovations in the Ediacaran Period and later in the Cambrian Period. The increase in size, number of comb rows, and complexity of neurological structures shows a not yet complete co-option of the cellular elements of proliferation, mechanical socialization, and cell differentiation and a transition toward complete co-option and definitive self-organization of the first embryos.

**Concluding remarks and perspectives**

The embryo was born free, but it is chained in every sense. It was not a political society (Rousseau, 2002) but rather a biophysical multicellular socialization phenomenon propelled by Neoplasia in the environmental context of the Ediacaran ocean with properties acquired from unicellular Holozoa. The formation of the first animal cell created a physical framework capable of containing the Neoplastic force; therefore, this force was born free but was free to be controlled. Joseph Needham acknowledged "the controlling forces from which the cancerous growth has escaped" (Needham, 1936). Thus, animal organization was generated by a balance between Neoplasia and mechanical forces that shaped chromatin structure and allowed to establish a module and process compatible with the containment of this force to form an embryo. Finally, it should be clarified that i do not know how long the period of frank expansion of Neoplasia lasted, nor how many attempts were needed to form an embryo; i only know that the process stabilized. This idea agrees with the paleontological hypothesis of punctuated equilibrium proposed by Stephen Jay Gould and Niles Eldredge (Gould and Eldredge, 1977; Gould, 1989) and with phylostratigraphic analyses of cancer genes (Domazet-Lo¬o and Tautz, 2010). I can predict that, after stabilization, Neoplasia, as the main core of evolution, was more contained and became a disease (cancer), with occasional or recurrent events in all animal groups (Sparks, 1972; Kaiser, 1989; Rothschild, Witzke and Hershkovitz, 1999; Robert, 2010; Tascedda and Ottaviani, 2014; Aktipis *et al.*, 2015; Sinkovics, 2015; Albuquerque *et al.*, 2018).

From this perspective, i envision a change of course in the search to understand the control and cure of cancer. Cell invasion and migration are natural and necessary events to constitute an embryo. How was cell invasion controlled in the embryo? The clues lie in evolution and the embryonic context. A comparison of the NFM during evolution and its physical organization in chromatin may provide insights into cancer control and containment. Further clues may be provided by the biophysical and mechanical context that permeated cancer and the embryo. The multiple activities of the actin cytoskeleton are part of evolutionary spandrels and emergent properties of the embryo that are not encoded in the genetic material. But which embryonic model should be used for this investigation? I envision the partial abandonment of classical models of development and incursion into animal models that will help to understand the great developmental transitions and shaping of cancer. Ctenophora, Cnidaria, and Porifera are models that may constitute the first step in gaining answers for embryogenesis as well as for cancer, a disease for which we have little knowledge on the cure (Smithers, 1962) but that is ingrained in the foundations of animal construction.

**Acknowledgments**

We thank Dr. Jose Bastos, pathologist and oncologist, for his permanent contribution to my cancer research projects. The authors offer apologies to all researchers who were not mentioned in the article, given the need to establish priorities in the article's construction.

**Conflict of interest statement**

The authors declare that the research was conducted in the absence of any commercial or financial relationships that could be construed as a potential conflict of interest.

associated domains in eukaryotes', *bioRxiv*. Cold Spring Harbor Laboratory, p. 2021.10.08.463732. Available at: https://www.biorxiv.org/content/10.1101/2021.10.08.463732v1%0Ahttps://www.biorxiv.org/content/10.1101/2021.10.08.463732v1.abstract.

Liu, D. *et al.* (2022) 'IGF2BP2 promotes gastric cancer progression by regulating the IGF1R-RhoA-ROCK signaling pathway', *Cellular Signalling*, 94, p. 110313. doi: 10.1016/j.cellsig.2022.110313.

Liu, X. *et al.* (2020) 'Matrix Metalloproteinases in Invertebrates', *Protein & Peptide Letters*, 27(11), pp. 1068–1081. doi: 10.2174/0929866527666200429110945.

Loganathan, R. *et al.* (2012) 'Spatial Anisotropies and Temporal Fluctuations in Extracellular Matrix Network Texture during Early Embryogenesis', *PLoS ONE*, 7(5), p. e38266. doi: 10.1371/journal.pone.0038266.

van Loon, A. P. *et al.* (2020) 'Cortical contraction drives the 3D patterning of epithelial cell surfaces', *Journal of Cell Biology*, 219(3), p. e201904144. doi: 10.1083/jcb.201904144.

Mackie, G. O., Mills, C. E. and Singla, C. L. (1988) 'Structure and function of the prehensile tentilla of Euplokamis (Ctenophora, Cydippida)', *Zoomorphology*, 107(6), pp. 319–337. doi: 10.1007/BF00312216.

Maître, J.-L. *et al.* (2012) 'Adhesion Functions in Cell Sorting by Mechanically Coupling the Cortices of Adhering Cells', *Science*, 338(6104), pp. 253–256. doi: 10.1126/science.1225399.

Maître, J.-L. *et al.* (2015) 'Pulsatile cell-autonomous contractility drives compaction in the mouse embryo', *Nature Cell Biology*, 17(7), pp. 849–855. doi: 10.1038/ncb3185.

Mañes, S. *et al.* (1997) 'Identification of Insulin-like Growth Factor-binding Protein-1 as a Potential Physiological Substrate for Human Stromelysin-3', *Journal of Biological Chemistry*, 272(41), pp. 25706–25712. doi: 10.1074/jbc.272.41.25706.

Mangoud, N. O. M. *et al.* (2021) 'Chitinase 3-like-1, Tolloid-like protein 1, and intergenic
70

Wirtz, D., Konstantopoulos, K. and Searson, P. C. (2011) 'The physics of cancer: The role of physical interactions and mechanical forces in metastasis', *Nature Reviews Cancer*, 11(7), pp. 512–522. doi: 10.1038/nrc3080.

Wolpert, L. (1969) 'Positional information and the spatial pattern of cellular differentiation', *Journal of Theoretical Biology*, 25(1), pp. 1–47. doi: 10.1016/S0022-5193(69)80016-0.

Wolpert, L. (1971) 'Positional Information and Pattern Formation', *Current topics in developmental biology*, 6, pp. 183–224. doi: 10.1016/S0070-2153(08)60641-9.

Wolpert, L. and Gustafson, T. (1961) 'Studies on the cellular basis of morphogenesis of the sea urchin embryo: The formation of the blastula', *Experimental Cell Research*, 25(2), pp. 374–382. doi: https://doi.org/10.1016/0014-4827(61)90287-7.

Wood, W. *et al.* (2002) 'Wound healing recapitulates morphogenesis in Drosophila embryos', *Nature Cell Biology*, 4(11), pp. 907–912. doi: 10.1038/ncb875.

Wozniak, M. A. and Chen, C. S. (2009) 'Mechanotransduction in development: A growing role for contractility', *Nature Reviews Molecular Cell Biology*, 10(1), pp. 34–43. doi: 10.1038/nrm2592.

Wu, X. *et al.* (2014) 'miR-194 suppresses metastasis of non-small cell lung cancer through regulating expression of BMP1 and p27kip1', *Oncogene*, 33(12), pp. 1506–1514. doi: 10.1038/onc.2013.108.

Wymeersch, F. J., Wilson, V. and Tsakiridis, A. (2021) 'Understanding axial progenitor biology in vivo and in vitro', *Development*, 148(4), p. dev180612. doi: 10.1242/dev.180612.

Xiao, W. *et al.* (2020) 'Overexpression of BMP1 reflects poor prognosis in clear cell renal cell carcinoma', *Cancer gene therapy*. 2019/06/03, 27(5), pp. 330–340. doi: 10.1038/s41417-019-0107-9.

Xiong, F. *et al.* (2014) 'Interplay of Cell Shape and Division Orientation Promotes Robust Morphogenesis of Developing Epithelia', *Cell*, 159(2), pp. 415–427. doi:

**Figures**

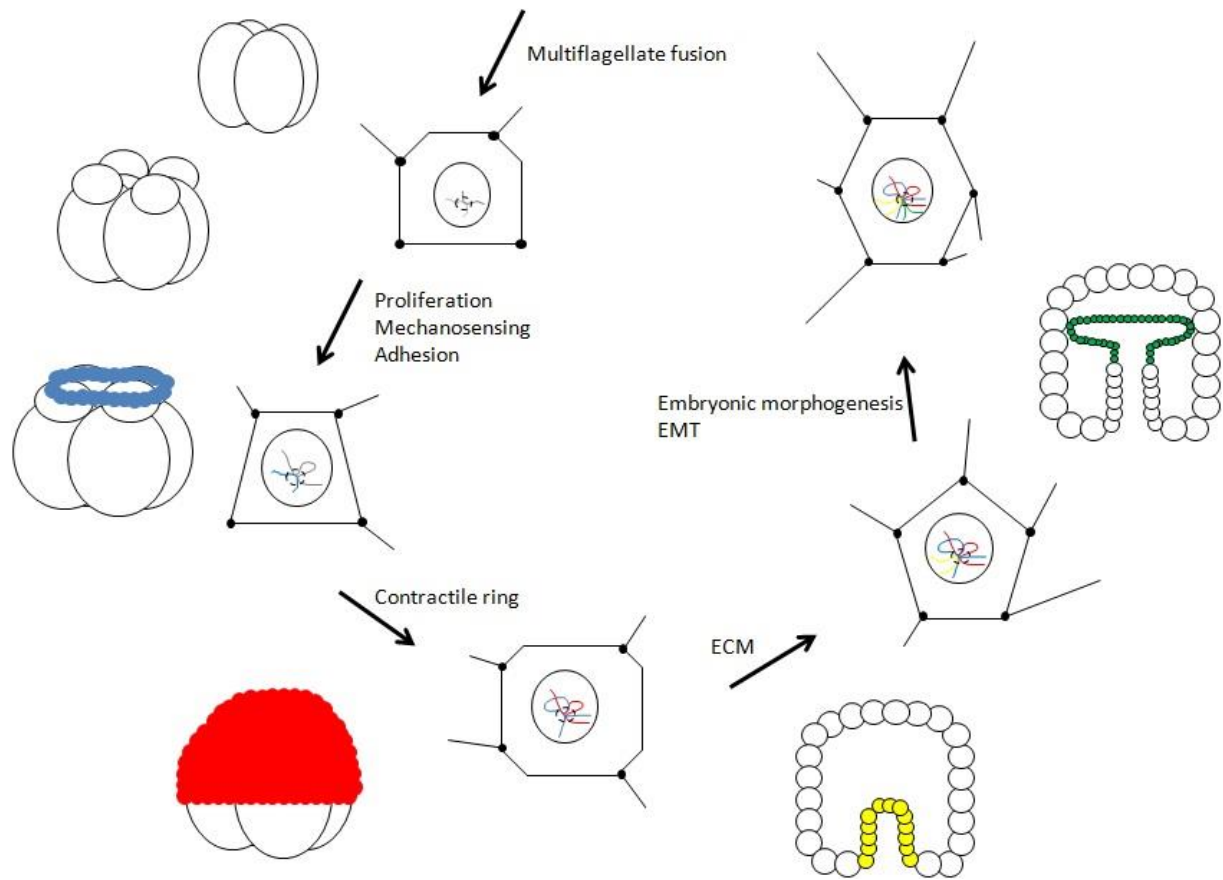

Figure 1. General scheme of the neoplastic functional module. Chromosomal domains were associated topologically by the physical impact of embryo construction. Epiboly and the extracellular matrix are two exaptations fundamental for embryo construction. The projected points and lines of the geometric figures represent the impact of embryonic morphogenesis on the cell nucleus. The cells depicted in the figure are multipotent cells that receive most of the biophysical impact promoting embryo patterning. Neoplasia is the driving force of embryo formation. Multiflagellate fusion is the initial event of embryogenesis, as will be discussed in part III of the hypothesis. The disease cancer is imbued in embryo construction and masked by its organization. The last cell of the scheme is one of the two cells of the hermaphroditic germ line of the first embryo, which harbors the neoplastic functional module and mechanical memory, elements that contribute to the reconstruction of the process in the following generation. Cell differentiation is revealed throughout the process as a consequence of morphogenesis and biophysical impacts on the first embryo. Emergence of animal

phylogeny, cancer, and the first embryo is implicit in the events represented in this diagram. The two great biophysical revolutions contributed significantly to embryonic morphogenesis and, consequently, to cell differentiation. EMT, Epithelial–Mesenchymal Transition; ECM, Extracellular Matrix or Basement Membrane.

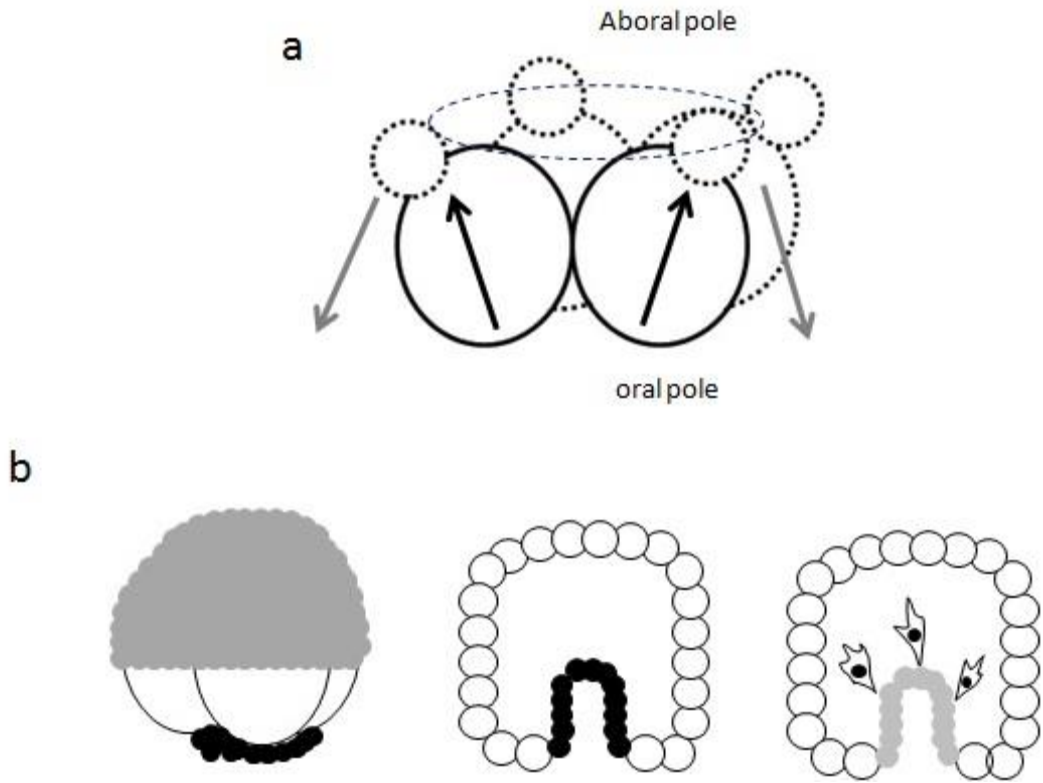

Figure 2. Forces generated by epiboly contribute to gastrulation in the first embryo. (a) Cells undergoing epiboly are believed to produce a mechanical force in the oral direction (gray arrow), whereas macromeres would produce a reciprocal force (mechanoreciprocity) in the aboral direction. As the epibolic process unfolds, the extracellular matrix is assembled and its rigidity contributes to the consolidation of gastrulation. (b) As a consequence of epiboly, an increase, of sufficient magnitude and duration, in the tension exerted on the cytoskeleton of macromeres promotes cell proliferation, noticeable as an increase in micromere formation, disturbances in tissue organization (invagination), epithelial–mesenchymal transition, and cell migratory capacity. The model of this figure was inspired by the results of Paszek et al.

(2005).